\documentclass[journal, 10pt]{IEEEtran}
\makeatletter
\def\ps@headings{%
\def\@oddhead{\mbox{}\scriptsize\rightmark \hfil \thepage}%
\def\@evenhead{\scriptsize\thepage \hfil \leftmark\mbox{}}
\def\@oddfoot{}%
\def\@evenfoot{}}
\makeatother
\usepackage{cite}

\ifCLASSINFOpdf
  \usepackage[pdftex]{graphicx}
  \DeclareGraphicsExtensions{.pdf,.jpeg,.png}
\else
  \usepackage[dvips]{graphicx}
\fi

\usepackage[cmex10]{amsmath}
\usepackage{amssymb}
\usepackage{verbatim}

\usepackage{url}

\usepackage{leading}
\leading{12pt}

\usepackage{ifthen}
\def\Context{TR}
\DeclareMathOperator*{\argmax}{arg\,max}
\makeatletter
\def\blfootnote{\xdef\@thefnmark{}\@footnotetext}
\makeatother

\begin{document}
\newtheorem{theorem}{Theorem}[section]
\newtheorem{lemma}{Lemma}[section]
\newtheorem{corollary}{Corollary}[section]
\newtheorem{proposition}{Proposition}[section]
\newtheorem{definition}{Definition}[section]
\newtheorem{claim}{Claim}[section]
\newtheorem{remark}{Remark}[section]
\newtheorem{example}{Example}[section]

\def\bydef{\buildrel \triangle \over =}
\newcommand{\ignore}[1]{}
\newcommand{\bff}{{\bf f}}
\newcommand{\bfi}{{\bf f^i}}
\newcommand{\bfl}{{\bf f_l}}
\newcommand{\bJ}{{\bf J}}
\newcommand{\cG}{{\cal G}}
\newcommand{\cJ}{{\cal J}}
\newcommand{\cN}{{\cal N}}
\newcommand{\cI}{{\cal I}}
\newcommand{\cL}{{\cal L}}
\newcommand{\cV}{{\cal V}}
\newcommand{\cF}{{\cal F}}
\newcommand{\cS}{{\cal S}}
\newcommand{\hfl}{f_{\hat{l}}}
\newcommand{\hfli}{f_{\hat{l}} ^i}
\newcommand{\hflj}{f_{\hat{l}} ^j}
\newcommand{\fhl}{\hat{f_l}}
\newcommand{\fhli}{\hat{f_l} ^i}
\newcommand{\fhlj}{\hat{f_l} ^j}
\newcommand{\fl}{f_l}
\newcommand{\fli}{f_l ^i}
\newcommand{\flj}{f_l ^j}
\newcommand{\ftl}{\tilde{f_l}}
\newcommand{\ftli}{\tilde{f_l} ^i}
\newcommand{\fuv}{f_{uv}}
\newcommand{\fuvi}{f_{uv} ^i}
\newcommand{\fuvj}{f_{uv} ^j}
\newcommand{\fvui}{f_{vu} ^i}
\newcommand{\fvuj}{f_{vu} ^j}
\newcommand{\hl}{\hat{l}}
\newcommand{\hbf}{\hat{\mbox{\bf f}}}
\newcommand{\hli}{\hat{\lambda}^i}
\newcommand{\la}{\lambda}
\newcommand{\li}{\lambda^i}
\newcommand{\hlj}{\hat{\lambda}^j}
\newcommand{\lj}{\lambda^j}

\newcommand{\Jsys}{J_{sys}}
\newcommand{\Jli}{J_l ^i}
\newcommand{\Khl}{K_{\hat l}}
\newcommand{\Khli}{K_{\hat l} ^i}
\newcommand{\Khlj}{K_{\hat l} ^j}
\newcommand{\Kl}{K_l}
\newcommand{\Kli}{K_l ^i}
\newcommand{\Klj}{K_l ^j}
\newcommand{\Tl}{T_l}
\newcommand{\Thl}{T_{\hat{l}}}
\title{How Good is Bargained Routing?}

\author{\IEEEauthorblockN{Gideon Blocq}
and
\IEEEauthorblockN{Ariel Orda}}

\maketitle

\thispagestyle{headings}
\pagestyle{headings}

\blfootnote{G. Blocq and A. Orda are with the Department of Electrical
Engineering, Technion, Haifa 32000, Israel (e-mails: gideon@tx.technion.ac.il,
ariel@ee.technion.ac.il).}

\begin{abstract}
In the context of networking, research has focused on {\em non-cooperative} games, where the {\em selfish agents} cannot reach a binding agreement on the way they would share the 
infrastructure. Many approaches have been proposed for mitigating the typically inefficient operating points. 
However, in a growing number of networking scenarios 
selfish agents are able to communicate and reach an agreement.
Hence, the degradation of performance should be considered at  
an operating point of a {\em cooperative game}.
Accordingly, our goal is to lay foundations for the application of
cooperative game theory to fundamental problems in networking.
We explain our choice of the {\em Nash Bargaining Scheme (NBS)} as the
solution concept, and introduce the {\em Price of Selfishness (PoS)},
which considers the degradation of performance at the worst {\em NBS}.
We focus on the fundamental load balancing game of routing over parallel links.
First, we consider agents with identical performance objectives.
We show that, while the {\em PoA} here can be 
{\em large}, through bargaining, 
all agents, and the system, strictly improve their performance. 
Interestingly,
in a two-agent system
or 
when all agents have identical demands, 
we establish that  
they reach {\em social optimality}.
We then consider agents with different performance objectives and 
demonstrate that the {\em PoS} and {\em PoA} can be unbounded, yet
we explain why both 
measures are unsuitable. Accordingly, we introduce the {\em Price of Heterogeneity (PoH)}, as an extension of the {\em PoA}. 
We establish an upper-bound on the {\em PoH}
and indicate its further motivation for bargaining. 
Finally, we discuss network design guidelines that follow from our findings.


\end{abstract}

\section{Introduction}

\subsection{Background and Motivation}
Traditional communication networks were designed and operated with systemwide optimization in mind.  
However, it has been recognized that systemwide optimization may be an
impractical paradigm for the control of modern networking configurations. Indeed, control decisions in large-scale networks are often made by
various agents independently, according to their individual
interests, and Game Theory~\cite{Myerson} provides the systematic framework to
study and understand their behavior.
To date, game theoretic models have been employed in virtually all
networking contexts. These include control tasks at the network layer,
such as flow control (e.g.,~\cite{korilis95existence}), 
routing (e.g.,~\cite{ORS93, AltmanBJS02, La02, Roughgarden:2002})
and multicasting as well as numerous studies on control tasks at the link and MAC layers.
Moreover, the application of Game Theory to communication networks has extended beyond 
control tasks. For example, several studies 
considered game theoretic scenarios in the context of 
the creation and evolution of the network topology. Others considered game theoretic scenarios at other layers, e.g., numerous studies in the context of network 
security (see~\cite{booksecurity} and references therein)
and a large body of work on peer-to-peer applications.  


Research to date
has mainly focused on {\em non-cooperative} networking games, where the {\em selfish} decision makers (i.e., the {\em players},
or {\em agents})
cannot communicate and reach a binding agreement on the way they would share the network infrastructure. Moreover,
the
main dynamics that were considered were {\em Best-Reply}, i.e., each player would observe the present state of the
network and react to it in a self-optimizing manner. Accordingly, the operating points of such systems
were taken to be some equilibria of the underlying non-cooperative game, most notably, {\em Nash equilibria}.

Such equilibria are inherently inefficient~\cite{Debreu1952} and, in
general, exhibit suboptimal network performance. As a result,
the question of how bad the quality of a Nash
Equilibrium is with respect to a centrally enforced optimum has
received considerable attention e.g.,~\cite{Koutsoupias99,Roughgarden:2002,EKL2003}. 
In order to quantify this inefficiency, several conceptual measures
have been proposed in the literature. 
In particular, the \emph{Price of Anarchy (PoA)}~\cite{Koutsoupias99},
defined as the ratio between the system (social) performance at a (worst) Nash Equilibrium
and the 
corresponding optimal system performance, has become the {\em de facto} benchmark for measuring the 
performance of non-cooperative networking games.

It has been repeatedly observed that the value of the 
{\em PoA} is typically large, often unbounded and many approaches have been proposed 
for mitigating this problem. These include schemes for resource provisioning~\cite{Shenker95},
{\em Stackelberg} strategies for controlling part of the 
traffic~\cite{KorilisLO97,Roughgarden:01}, 
incentive schemes for cooperation or cost sharing among mobile terminals~\cite{AltmanBKMM04, JS10},
schemes for choosing the initial configuration~\cite{stability_paper, ChekuriCLNO07},
schemes for exercising limited control on the game dynamics\cite{ChekuriCLNO07} and numerous proposals for pricing mechanisms. 
Some studies also considered players to be ``partially altruistic''~\cite{PAAA_WiOpt:2010}.
Nevertheless, in all the above studies, the standing assumption has been that the network agents play a {\em non-cooperative}
game. 

However, there is a growing number of networking scenarios where, while there is competition among 
self-optimizing agents (i.e., a ``game'' among ``selfish players''), there is also a possibility for these
agents to communicate, negotiate and {\em reach a binding agreement}. Indeed, in many scenarios
the competition is among business organizations, which can, and often do, reach agreements (e.g., SLAs) on the way
that they provide, consume or share the network resources. The proper framework for analyzing such
settings is that of {\em Cooperative Game Theory}\cite{Myerson}. Such a paradigm transfer, from non-cooperative 
to cooperative
games, calls to revisit fundamental concepts. Indeed, the operating point of the network is not an equilibrium of a non-cooperative game, but rather a solution concept of a cooperative
game. Accordingly, the performance degradation of such systems should be considered at the new operating
points. It is also important to note that the (typically high, often unbounded) Price of Anarchy is a 
price that is paid {\em not only due} to the selfish nature of the
decision makers, {\em but also due} to their inability to cooperate; when the latter becomes possible, we need
a measure that accounts for the price that is paid {\em solely} due to the selfishness of the network agents.

The goal of this paper is to lay foundations for the application of 
Cooperative Game Theory to fundamental problems in networking. 
We focus on the Nash Bargaining Scheme (NBS)~\cite{Myerson} as the
solution concept for cooperative networking games. As shall be discussed, per bargaining problem, 
the existence and uniqueness of a solution to the NBS is guaranteed under mild conditions. Accordingly, we introduce  a novel concept for 
measuring the effect of cooperation, termed as the {\em Price of Selfishness (PoS)}. Taken as the 
ratio between the system (social) performance under the (worst) {\em Nash Bargaining Scheme} and the 
corresponding optimal system performance, the {\em PoS} 
quantifies the loss incurred by agents solely due to their selfish behavior. 

\subsection{Previous Work}

Cooperative game theoretic models and solution concepts (and the NBS in particular) have been considered in the context of networking by a few studies. 
In \cite{PC06}, the NBS is used to distribute jobs fairly among servers, while in \cite{TAG06} it is used to allocate bandwidth fairly among users. 
In \cite{JZRM09}, the NBS is implemented to improve the fairness and efficiency of traffic engineering and server selection. 
In \cite{AEMN11}, network formation is addressed through the use of cooperative game theoretic tools such as the {\em Shapley Value}\cite{Myerson} and the NBS. 
There, it is numerically shown that the NBS permits to allocate costs fairly to users within a reasonable computation time. 
In \cite{MJWS_INFOCOM:2010}, the NBS is used to calculate the subgame perfect equilibrium and provide upper-bounds for the {\em PoA}.
In \cite{Gibbens:2008}, coalition games and the Shapley Value are used to maximize the utility framework for routing and flow control in ad-hoc networks. 
In \cite{HanawalA13} a modified version of the NBS is used to set fair prices between ISP and CP's in a nonneutral network.

Cooperative Game Theory and the NBS in particular, have also been used in spectrum sharing~\cite{SDHM2007}, where nodes in a multi-hop wireless
network need to agree on a fair allocation of the spectrum. In \cite{WU-BBS-SECON-09}, 
a bargain-based mechanism is proposed for message passing in participatory sensing networks, in order to encourage cooperative message trading among the
selfish nodes.
In \cite{NWSH2010}, a coalition game model 
with a stable solution is proposed in order to investigate the performance gain of multiple communities
in delay tolerant networks. In \cite{LXWG2011}, the design of new coalition-based dynamics is investigated in the
context of cognitive radio networks. 
In \cite{Antoniou:2009}, a network synthesis game is studied,
in which individual access networks with insufficient resources form coalitions in order to satisfy service demands. 
There, the {\em Core}~\cite{Myerson} of the game is investigated for several payoff allocations among the players.
In \cite{SA11}, various cost allocation schemes are studied for players that have the option to join coalitions of multicast services in a wireless network.
We also note that there is a body of work on network bargaining games, e.g.,
\cite{BHIM10} and references therein; however, in those studies a "network"  describes some relations among general economic agents.

Most related to the present paper, a previous study~\cite{AndelmanFM07}  
proposed {\em the Strong 
Price of Anarchy (SPoA)} as a 
measure that considers the degradation of performance when some collaboration among the agents is possible. The {\em SPoA} is
defined similarly to the {\em PoA} but
considers only {\em strong} (rather than all) Nash 
equilibria;\footnote{A Strong Nash Equilibrium is a Nash Equilibrium in which 
no coalition, taking the actions of its complements as given, can cooperatively deviate in a way that benefits 
all of its members.} however, since such equilibria are 
not guaranteed to exist (in particular, we indicate they do not in the framework considered in this paper), 
it cannot provide a general benchmark that would be the 
``cooperative games counterpart'' of the Price of Anarchy.

\subsection{Our Contribution}

We concretize our study by considering the setting of routing in a ``parallel links'' network. Beyond being a basic
framework of routing, this is the {\em generic 
framework of load balancing} among servers in a network.  It has been 
the subject of numerous studies in the context of non-cooperative networking games, e.g.,
~\cite{ORS93, Koutsoupias99, Roughgarden:01, La02, harks_09},
to name a few. In particular, in~\cite{ORS93} 
it has been established that, under a non-cooperative routing game and under some standard modeling 
assumptions, the system has a unique Nash Equilibrium.  
We begin by considering the case where $N$ agents aim 
at optimizing the same type of performance objective, e.g., each tries to minimize its traffic delay.
This is the classic setting on which the literature has focused, in particular whenever considering 
the Price of Anarchy.
We demonstrate that, in the considered routing game, 
the degradation of system performance at the Nash Equilibrium can be very large.
Yet, we establish that, for an interesting class of performance functions, the $N$-player bargaining problem related to our routing game, is {\em essential}, i.e., if the network agents are allowed to bargain and reach a binding 
agreement (while remaining selfish), each agent is guaranteed to {\em strictly} improve its performance.
Consequently, the {\em PoS} is {\em strictly} smaller than the {\em PoA}. 
Interestingly, when $N=2$ or when all $N$ agents have identical demands, 
the {\em PoS} is equal 
to $1$; that is, by letting the agents bargain, no loss in system performance is 
incurred due to their selfishness. 
On the other hand we provide an example where $N>2$ and show that $1<PoS<PoA$. 
For this case it remains an open question how to tighten the bounds on the {\em PoS}.

We then extend our study to address the case
where agents consider vastly different (``heterogeneous'') performance objectives
and demonstrate that the corresponding $N$-player bargaining game is {\em not necessarily essential}. We then show that the 
{\em PoS}, and also the {\em PoA}, can be unbounded. 
However,
we explain why both measures may be unsuitable for such heterogeneous scenarios. Accordingly, we introduce
an additional measure, termed the {\em Price of Heterogeneity (PoH)}, and  indicate that it
is a proper extension of the {\em PoA} for the heterogeneous setting. We establish an upper-bound on the {\em PoH} for 
a quite general class of (heterogeneous) performance objectives, and indicate that it provides incentives for bargaining also
in this, more general case.
Finally, we discuss some network design guidelines that follow from our findings.  

The main contributions of this study can be summarized as follows:
\begin{itemize}
\item We introduce the {\em Price of Selfishness} as a figure of merit for network (or system) performance under a cooperative game. 
\item We establish that, in the game of routing over parallel links
and for interesting classes of ``homogeneous'' performance objectives,
the $N$-player bargaining problem is {\em essential}, i.e., all agents {\em strictly} improve their performance, and the {\em PoS} is {\em strictly} smaller than the potentially large {\em PoA}. Moreover, when $N=2$ or when all $N$ agents have identical demands, the {\em PoS} is equal to $1$. 
\item We indicate that, in the wider case of ``heterogeneous'' performance objectives, the $N$-player bargaining problem is not necessarily essential and we show that the {\em PoS} can 
be arbitrarily large. 
\item We introduce the {\em Price of Heterogeneity (PoH)} as a proper extension of the {\em PoA} for the 
heterogeneous case and establish an upper-bound on the {\em PoH}.
\end{itemize}

The rest of this paper is organized as follows.
In section~\ref{sec:model} we formulate the model and
terminology. In section~\ref{sec:PoS_1} we consider the classic case of homogeneous performance objectives.
The heterogeneous case is treated in section~\ref{sec:PoH}. Finally, conclusions are presented in Section~\ref{sec:conclusion}.
\ifthenelse{\equal{\Context}{PAPER}}{
Due to space limits, some proofs and technical details are omitted
from this version and can be found (online) in~\cite{BO11}.
}
{
}
\section{Model and Game Theoretic Formulations}\label{sec:model}

\subsection{Model}
\label{model_a}

\begin{figure}[h!]
  \centering
    \includegraphics[width=0.5\textwidth]{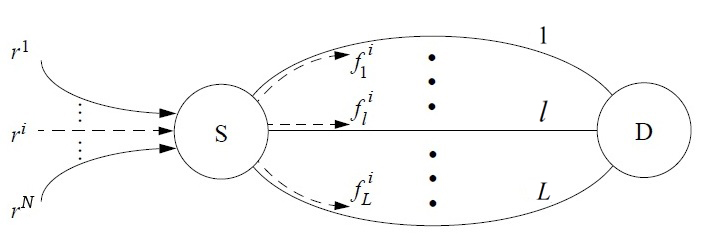}
  \caption{Parallel Links Model}
\label{fig:par_links}
\end{figure}

Following \cite{ORS93}, we are given a set $\cN = \{ 1,2, \ldots , N \}$ of
selfish ``users'' (or, ``players'', ``agents''), which share a set of
parallel ``links'' (e.g., communication links, servers, etc.) $\cL = \{ 1,2, \ldots ,L \}$, interconnecting a
common source node to a common destination node. See Figure \ref{fig:par_links}.
Let $c_l$ be the capacity of link $l$.
Each user  $i \in \cN$
has a traffic demand $r^i$.
A user ships its demand by splitting it over the links
$\cL$, i.e., user $i$ decides what fraction of
$r^i$ should be sent through each link.  We denote by $\fli$, the flow 
of user $i \in \cN$ on link $l \in \cL$. Thus, user $i$ can fix any
value for $\fli$, as long as $\fli \geq 0$ (non-negativity constraint)
and $\sum_{l \in \cL} \fli = r^i$ (demand constraint). Denote the total demand of all the users by $R$, i.e., $R = \sum_{i \in \cN} r^i$. We assume that the system of parallel links can accommodate the total demand, i.e., we only consider capacity configurations $\mathbf{c}=[c_1 \ldots c_L]$ for which $\sum_l c_l > R$.
Turning our attention to a link $l \in \cL$, let $\fl$ be the total flow
on that link i.e., $\fl = \sum_{i \in \cN} \fli$; also, denote by $\bfl$
the vector of all user flows on link $l \in \cL$, i.e., $\bfl = ( f_l ^1
, f_l ^2 , \ldots , f_l ^N )$.
The {\it routing strategy} of user $i$, $\bfi$, is the vector $\bfi = (
f_1^i , f_2^i , \ldots , f_L^i )$. The {\it (routing) strategy profile} $\bff$ is
the vector of all user routing strategies, 
$\bff = ( {\bf f^1} , {\bf f^2} , \ldots , {\bf f^N} )$. We say that a user's 
routing strategy is {\it feasible} if its components obey the nonnegativity
and demand constraints and we denote by $\bf F^i$ the set of all
feasible $\bfi$'s. Similarly, a routing strategy profile is feasible if
it is composed of feasible routing strategies and we denote by
$\bf{F}$ the set of all feasible $\bff$'s. 

The performance measure of a user $i \in \cN$ is given by a cost
function $J^i ( \bff )$. The aim of each user is to minimize its cost.
As in~\cite{ORS93}, the following standard assumptions on the cost function $J^i$ of each user
are imposed:
\begin{list}{S\theenumi}{\usecounter{enumi}}
\item $J^i$ is the sum of link cost functions i.e., \\
$J^i ( \bff )= \sum_{l \in \cL} \Jli ( \bfl )$. 
\item $\Jli$ is a function of two arguments, namely user $i$'s flow on link
$l$ and the total flow on that link. In other words: $\Jli ( \bfl) = J_l
^i ( \fli , \fl)$. 
\item $J_l ^i$ is increasing in each of its two arguments.
\item $\Jli : [0, \infty )^2 \rightarrow [0, \infty )$, a continuous function.
\item $\Jli$ is convex in $\fli$.
\item $\Jli$ is continuously
differentiable in $\fli$. 
\item Note that $\frac{\partial \Jli}{\partial \fli} = \frac{\partial \Jli}{\partial \fli} ( \fli , \fl )$, i.e., it is a function
of two arguments. We assume that whenever $\Jli$ is finite,
$\frac{\partial \Jli}{\partial \fli} ( \fli , \fl )$
is strictly increasing in each of the two arguments.
\end{list}

Cost functions that comply with the above assumptions shall be referred to as
{\it Standard}. An $N$-tuple of positive values \\ $\bJ = (J^1, J^2, \ldots , J^N)$ is said to be
a {\em feasible cost vector} if there is a feasible (routing) strategy profile $\bff \in \bf {F}$ such that, for all $1 \leq i \leq N$,
$J^i = J^i(\bff)$. 

An important class of problems is when users are interested in the same performance measure,
e.g., delay (i.e., each user aims at optimizing the delay of {\em its} traffic). In fact, much of the current
literature on networking games has focused on this class, e.g., ~\cite{Korilis97capacityallocation, Koutsoupias99,AltmanBJS02, La02, Roughgarden:2002, harks_09, BhaskarFHH09}.
In this case, the performance of a link $l$ is manifested through some
function $\Tl ( \fl )$, which measures the cost per unit of flow on the link,
and depends on the link's total flow. For example, $\Tl$ may be the delay of link
$l$. Specifically, we consider users whose cost functions assume the following, ``homogeneous'', form:
\begin{list}{H\theenumi}{\usecounter{enumi}}
\item $\Jli ( \fli , \fl ) = \fli \cdot \Tl ( \fl )$.
\item $\Tl : [0, \infty ) \rightarrow [0, \infty )$.
\item $\Tl ( \fl )$ is strictly increasing and convex.
\item $\Tl ( \fl )$ is continuously differentiable.
\item $\Tl ( \fl ) = \begin{cases}  T(c_l-f_l) & \text{if}~f_l < c_l 
\\ \infty & \text{if}~f_l\geq c_l, 
\end{cases}$ 
\end{list}
where the function $T(\cdot)$ is independent of the link entity, but it is a function of the residual capacity $c_l-f_l$. Moreover, $T(c_l-f_l)$ is strictly increasing in $f_l$. 
Cost functions that comply with the above assumptions shall be referred to as
{\it homogeneous}. Note that homogeneous functions are necessarily standard.
We note that, for homogeneous functions, we have
$$
\frac{\partial \Jli}{\partial \fli} (\fli, \fl) = \fli \cdot \Tl ' (\fl)  + \Tl (\fl)
$$
where $\Tl ' = \frac{d \Tl}{d \fl}$. 
Note also that, if $\Tl ( \fl )$ is the average delay per unit of flow,
then the corresponding homogeneous cost function is the widely used total delay
function (in our case, per user); by dividing the latter by the traffic
demand we obtain the traffic's average delay. 
Finally, we note that the technical 
Assumption H5 holds for interesting classes of cost functions, such as link delays under the $M/M/1$ queueing model, where
$T_l(f_l) = \frac{1}{C_l-f_l}$. 
As a result of  Assumption H5, we get the following lemma.
\begin{lemma}
\label{homogeneous_0}
For any two links $l,n$ it holds that $T_l(f_l) \leq T_n(f_n)$ if and only if $T_l'(f_l) \leq T_n'(f_n)$.
\end{lemma}
\begin{IEEEproof}
\ifthenelse{\equal{\Context}{PAPER}}
{See \cite{BO11}.
}
{
\noindent \textbf{Necessary Condition}:

First consider $c_l \geq c_n$ and denote $\delta \equiv c_l-c_n$.
From Assumption {\em H5} we get, 
$$T_l(f_l) \leq T_n(f_n) = T(c_n-f_n) = T(c_l-\delta-f_n) = T_l(f_n + \delta).$$
It follows from Assumption {\em H3} that $f_l \leq f_n + \delta$ and that $T_l'(f_l) \leq T_l'(f_n+\delta)$. Again, from Assumption {\em H5} we have, 
$T_l'(f_l) \leq T_l'(f_n+\delta) = T_n'(f_n)$.

Now consider $c_l < c_n$ and denote $\Delta \equiv c_n-c_l$.
From Assumption {\em H5} we get, 
$$T_l(f_l) = T(c_l-f_l) = T(c_n-\Delta-f_l) = T_n(f_l + \Delta) \leq T_n(f_n).$$
It follows that $f_l + \Delta \leq f_n$ and $T_l'(f_l) = T_n'(f_l + \Delta) \leq T_n'(f_n)$.

\noindent \textbf{Sufficient Condition}
Due to Assumptions {\em H3} and {\em H5}, the proof follows straight from the necessary condition's proof when switching $T(\cdot)$ with $T'(\cdot)$.
} 
\end{IEEEproof}
We order the links such that $\forall l<n, c_l \geq c_n$, i.e., $\forall l<n$, $\forall f$, $T_l(f) \leq T_n(f)$. 
Hence, Assumption H5 effectively implies a ``quality'' ordering of the links.
In the rest of the section, cost functions shall be assumed to be just ``standard'', unless explicitly referred to as ``homogeneous''.

\subsection{Game Theoretic Formulation}
\label{gtf}
We distinguish between two cases, namely {\em noncooperative} and {\em cooperative} game scenarios, as follows.
\subsubsection{Noncooperative Routing (Load Balancing) Game}

In this case, the standard solution concept is the {\em Nash Equilibrium}\cite{Myerson}, i.e., 
a routing strategy profile such that no user finds it beneficial to change its flow
on any link.
Formally, a feasible routing strategy profile
$\bf{\hat{f}} = ( \bf{\hat{f}^1} , \bf{\hat{f}^2} , \ldots ,
\bf{\hat{f}^N} )$ is a {\em Nash Equilibrium Point (NEP)} if, for all $i \in
\cN$,
the following condition holds:
\begin{align} \label{eq:e31}
J^i ( {\bf\hat{f}} ) &= 
J^i ( {\bf\hat{f}^1} , \ldots , {\bf\hat{f}^{i-1}} ,
{\bf\hat{f}^i}, {\bf\hat{f}^{i+1}}, \ldots , {\bf\hat{f}^N} )  \\ \nonumber
&= \min_{\bfi \in {\bf F^i}} J^i ( {\bf\hat{f}^1} ,
\ldots , {\bf\hat{f}^{i-1}} ,
{\bf f^i}, {\bf\hat{f}^{i+1}}, \ldots , {\bf\hat{f}^N}).
\end{align}

It follows from our assumptions on standard cost functions that the minimization in
(\ref{eq:e31}) is equivalent to the following {\em Karush-Kuhn-Tucker (KKT)} conditions:
for every $i \in \cN$ there exist a (Lagrange multiplier) $\lambda ^i$
such that, for every link $l \in \cL$,
\begin {equation} \label{kut1}
\fli >0 \rightarrow \frac{\partial \Jli}{\partial \fli} ( \bfl ) = \lambda ^i
\end {equation} \begin{equation*}
\fli =0 \rightarrow \frac{\partial \Jli}{\partial \fli} ( \bfl ) \geq \lambda ^i.
\end{equation*}
The KKT conditions as stated above constitute
necessary and sufficient conditions for a
feasible routing strategy profile to be an NEP. In \cite{ORS93}, the following has been established:

\begin{theorem} \label{the:t1}
In a network of parallel links (as defined above), where the cost function of each user is
standard, there exists a routing strategy profile that is an NEP and it is unique.\footnote{On the other hand, it is easy to
verify that, in general, the unique NEP of the routing game is not a Strong Nash Equilibrium, due to the lack of Pareto optimality. Hence the 
Strong Nash Equilibrium cannot serve as a solution concept for the considered cooperative routing game.} 
\end{theorem}

For $1 \leq i \leq N$, denote by ${\bf \hat{J}} = (\hat{J}^1, \ldots , \hat{J}^N)$ the cost vector at the (unique) Nash Equilibrium.

\subsubsection{Cooperative Game}
The {\em Nash Bargaining Scheme} (or, {\em Nash Bargaining Solution}) is a main solution concept in Cooperative Game Theory~\cite{Myerson}.
Due to its appealing properties, such as its existence and uniqueness for each bargaining problem (under mild conditions), it has
been widely applied to cooperative scenarios.
Accordingly, we adopt it as our solution concept for the cooperative version of our routing game (and, more generally, for networking games).

Informally, a ``bargaining scheme'' proposes a cost vector that induces a strategy profile that the players agree to play. In the context of the Nash Bargaining Scheme, a standard assumption (see \cite{Myerson})
is that the agreement may consist of choosing between a finite number of strategy profiles, according to given (agreed) probabilities.
Then, the players would play the game according to the chosen (pure, in our case) strategy profile. This is the standard way to
cope with a certain convexity requirement, as explained in the following.

Formally, a bargaining scheme proposes a cost vector $\bf{\tilde g}$ that can be represented as the convex combination of some
feasible cost vectors, i.e. ${\bf{\tilde{g}}} = \sum_{m = 1}^M \tilde{p}_m \cdot J(\bf{\tilde{f}(m)})$, for some \\
$0 < \tilde{p}_m \leq 1, \sum_{m = 1}^M \tilde{p}_m = 1$ and $\bf{\tilde f(m)} \in \mathbf{F}$ for $m = 1 \dots M$, where $M$ is some finite number. $\tilde{\bf{p}} = [\tilde{p}_1, \ldots, \tilde{p}_M]$ is termed the {\em bargained
probability vector} and $\bf{\tilde f(m)}$ for $m = 1 \dots M$, are termed {\em bargained strategy profiles} (if their choice is not
unique we pick them arbitrarily). The set of all such $\bf{\tilde{g}}$ is denoted by ${\cG}$.
A {\em bargaining game} is defined by a {\em set of (bargainable) costs} $\cG$, as defined above, and by
a {\em disagreement point} $\textbf{v}$.
The disagreement point is a (feasible) cost vector that corresponds to
the costs that would be paid by the players if they do not reach an agreement.
As is usually done, we consider it to be the cost vector that
corresponds to the (unique) NEP, i.e., $\textbf{v} = {\bf \hat{J}}$, \cite{Myerson}.
Then,  an {\em $N$-player bargaining problem} is defined as follows:

\begin{definition}\label{def:begin_def}
An {\em $N$-player bargaining problem} consists of a pair $({\cG},\mathbf{\hat{J}})$, where ${\cG}$ is a closed convex subset of
$\mathbf{R}^N$,
$\mathbf{\hat{J}}$ is a vector in $\mathbf{R}^N$ and the
set {${\cG} \bigcap \{(g^1, \ldots, g^N) \mid g^i \leq {\hat{J}}^i \hspace{2 mm} \forall i \}$} is nonempty and bounded.
An $N$-player bargaining problem is {\em essential} if and only if there exists at least one cost vector $\bf{ g}$ for which ${g}^i < \hat{J}^i,~\forall i \in \cN$.
\end{definition}
It is easy to verify that the routing game considered in this study meets the mathematical requirements of an $N$-player bargaining problem. However, it is not clear this problem is also {\em essential}, i.e., that every user stands to strictly lower its cost through bargaining. 
In Sections \ref{sec:PoS_1} and \ref{sec:PoH} we will show that, under homogeneous costs, the $N$-player bargaining problem related to our routing game, is essential, while under standard costs this may not be the case.

We note that, in general, the convexity requirement imposed on ${\cG}$ in Definition \ref{def:begin_def} is necessary for obtaining
the structural
result of the following Theorem~\ref{the:t2}. Allowing players to bargain a random selection of the
strategy profile is the standard (and often only) way to obtain such convexity (see~\cite{Myerson}).
Nonetheless, we shall show that the results of this study can also be obtained when considering the more restricted (and more practically appealing) 
case in which no randomization is allowed and one bargained strategy profile should be chosen.

Given an $N$-player bargaining problem, a {\em Nash Bargaining Scheme (NBS)} is a cost vector \\
${\bf\tilde{g}} =
{\bf\tilde{g}}({\cG},\mathbf{\hat{J}}) = (\tilde{g}^1, \ldots , \tilde{g}^N)$ that satisfies
the following axioms~\cite{Myerson,rat_behav}:
\begin{list}{N\theenumi}{\usecounter{enumi}}
\item {\em Individual Rationality}: for $1 \leq i \leq N$, $\tilde{g}^i \leq \hat{J}^i$; i.e., no player will
incur a higher expected cost than at the disagreement point. If $1 \leq i \leq N$, $\tilde{g}^i < \hat{J}^i$, we say that this axiom is strictly satisfied. 
\item {\em Pareto Optimality}: If $\exists ~g \in {\cG}$ such that, for some $i$, $g^i < \tilde{g}^i$
then for some $k$ holds that $g^k > \tilde{g}^k$;
i.e., there is no way to reduce the expected cost of a player without increasing the cost of another player.
\item {\em Symmetry}: For any two players $i,k$, if $\hat{J}^i = \hat{J}^k$ and
$\{(\ldots,g^{k-1},g^k,\ldots,g^{i-1},g^i,\ldots) \mid$ \\
$(\ldots,g^{k-1},g^i,\ldots,g^{i-1},g^k,\ldots) \in \cG\} = \cG,$
then $\tilde{g}^i = \tilde{g}^k$;
i.e., if the players are indistinguishable, then the agreement should not discriminate between them.\footnote{Note that players with the same cost functions
and traffic demands are indistinguishable in the sense of this axiom.}
\item {\em Invariance to Equivalent Payoff Representations}: For any numbers $a^i, b^i$ with $i = 1 \ldots N$ and $a^i > 0$
for all $i$, if
${\cG} = \{(a^1 \cdot g^1+b^1, \ldots a^N\cdot g^N+b^N) \mid (g^1,\ldots ,g^N) \in {\cG}\}$ and $\textbf{w} = (a^1 \cdot \hat{J}^1+b^1, \ldots a^N \cdot \hat{J}^N+b^N)$
then $\tilde{\textbf{g}}(\cG,\textbf{w}) =$ $(a^1\cdot \tilde{g}^1({\cG},\mathbf{\hat{J}}) + b^1,$$ \ldots ,a^N \cdot \tilde{g}^N({\cG},\mathbf{\hat{J}}) + b^N)$;
i.e, a linear transformation of the utility function (being a transformation that maintains the some
ordering over preferences) should not
alter the outcome of the bargaining process.
\item {\em Independence of Irrelevant Alternatives}: For any closed convex set $\cS$, if
$\cS \subseteq {\cG}$ and $\tilde{\textbf{g}} \in \cal{S}$,
then $\tilde{\textbf{g}}(\cal{G},\mathbf{\hat{J}}) =$ $\tilde{\textbf{g}}$ $({\cS},\mathbf{\hat{J}})$;
i.e., removal of ``uninteresting'' strategy profiles should not alter the outcome of the bargaining process.
\end{list}

In \cite{Myerson}, the following has been established:

\begin{theorem} \label{the:t2}
For every $N$-player bargaining problem there exists a unique solution (i.e., NBS) ${\bf \tilde{g}}$ that satisfies all five axioms {\em (N1)-(N5)},
and it is provided by:
\begin{equation} \label{eq:NBS}
{\bf \tilde{g}} = \argmax_{\footnotesize{{\bf g} \in {\cG}, \forall i~g^i \leq \hat{J}^i}} \prod_{i=1}^N \left( \hat{J}^i - g^i \right) .
\end{equation}
\end{theorem}
Therefore, under mild conditions, a cooperative game always admits a solution
in the form of an NBS, which is unique in terms of the proposed cost vector.

\subsection{System Optimization}
As commonly assumed in the literature (e.g., \cite{KorilisLO95, KorilisLO97, Koutsoupias99, Roughgarden:01, La02, Roughgarden:2002, stability_paper, ChekuriCLNO07}), the welfare of a system is measured by the sum of the individual costs of the players, i.e.,
by a (``social'') cost function $\Jsys$ defined as $\Jsys = \sum_{i \in \cN} J^i$. We denote by $J^*_{sys}$ the optimal value of the system's cost, i.e., the minimal value of $\Jsys$ over all feasible
routing strategy profiles.
In addition, we denote by $\hat{J}_{sys}$ the value of the system cost at the (unique) NEP.
In the case of homogeneous cost functions, we have:
\begin{equation}
\label{social_cost}
\Jsys = \sum_{i \in \cN} J^i = \sum_{i \in \cN} \sum_{l \in \cL} f_l^i \cdot T_l(f_l) =
\sum_{l \in \cL} f_l \cdot T_l(f_l).
\end{equation}
For example, if
$T_l(f_l)$ stands for the link's delay, then $\Jsys$ corresponds to the total delay experienced by the system's traffic.
Note that, for homogeneous costs, $\Jsys$ depends only on the total flows on the links. Accordingly, for such costs, we denote by
$\bf{f}^*$ $= (f_l^*)_{l \in \cL}$ the optimal vector of link flows, i.e.,
\begin{equation}
J^*_{sys} = \sum_{l \in \cL} f_l^* \cdot T_l (f_l^*).
\end{equation}

Similarly, $\bf{\hat{f}}$ $= (\hat{f}_l)_{l \in \cL}$ is the vector of link flows at the NEP and
\begin{equation}
\hat{J}_{sys} = \sum_{l \in \cL} \fhl \cdot T_l (\fhl)
\end{equation}
is the system's cost at the NEP. Also, for an NBS with disagreement point $\mathbf{\hat{J}}$, bargained probability vector $\tilde{\bf{p}}$ and strategy profiles
$\bf{\tilde{f}(m)}$ for $m = 1 \dots M$,
$\bf{\tilde{f}(m)}$ $= (\tilde{f}_l (m))_{l \in \cL}$ are the corresponding vectors of link flows, and

\begin{equation}
\tilde{g}({\cG},\mathbf{\hat{J}})_{sys}  = \sum_{m = 1}^M \tilde{p}_m \cdot \left[ \sum_{l \in \cL} \tilde{f}_l (m) \cdot T_l (\tilde{f}_l (m)) \right]
\end{equation}

is the expected social cost at the NBS.

The following lemma establishes that, for homogeneous cost functions, both at the system optimum and at the Nash Equilibrium, the costs of
the links monotonically increase with the link index, i.e., ``better'' links bear lower costs.

\begin{lemma}
\label{homogeneous_1}
With homogeneous costs, $\forall l,1 \leq l < L$, the following hold: (i) $T_l(f_l^*) \leq T_{l+1}(f_{l+1}^*)$; and (ii) 
$T_l(\hat{f}_l) \leq T_{l+1}(\hat{f}_{l+1})$. 
\end{lemma}
\begin{IEEEproof}
\ifthenelse{\equal{\Context}{PAPER}}
{See \cite{BO11}.
}
{
Consider some arbitrary link $l, 1 \leq l \leq L-1$ at the system optimum. There are two possible cases: 

\begin{enumerate}
\item  $f_l^* \leq f_{l+1}^*$.
\item  $f_l^* > f_{l+1}^*$.
\end{enumerate} 

Consider Case 1. From Lemma \ref{homogeneous_0} it follows that \\
$T_l(f_l^*) \leq T_{l+1}(f_l^*) \leq  T_{l+1}(f_{l+1}^*)$, as required. 

Consider now Case 2. We have $f_l^* > f_{l+1}^* \geq 0$. The KKT conditions for system optimality induce:

\begin{equation}
\label{KKT_proof_1}
f_l^* T_l^{\prime}(f_l^*) + T_l(f_l^*) \leq f_{l+1}^* T_{l+1}^{\prime}(f_{l+1}^*) + T_{l+1}(f_{l+1}^*).  
\end{equation}

From (\ref{KKT_proof_1}) it follows that either $T_l(f_l^*) \leq T_{l+1}(f_{l+1}^*)$ or $T_l^{\prime}(f_l^*) \leq T_{l+1}^{\prime}(f_{l+1}^*)$. Due to Assumption {\em H5}, both statements are true, therefore $T_l(f_l^*) \leq T_{l+1}(f_{l+1}^*)$, as required.

Consider now the Nash Equilibrium and assume by contradiction that there exists a link $l, 0 \leq l \leq L-1$,
for which it holds that $T_l(\hat{f}_l)>T_{l+1}(\hat{f}_{l+1})$, hence, by Assumption {\em H5}, also
$T_l^\prime(\hat{f}_l)>T_{l+1}^\prime(\hat{f}_{l+1})$. If $\hat{f}_l=0$, it follows that $T_l(\hat{f}_l) = T_l(0)\leq T_l(\hat{f}_{l+1})\leq T_{l+1}(\hat{f}_{l+1})$, which contradicts $T_l(\hat{f}_l)>T_{l+1}(\hat{f}_{l+1})$. 

If $\hat{f}_l>0$, then $\forall i$ such that $\hat{f}_l^i>0$, the KKT condition (\ref{kut1}) induces:

\begin{equation}
\label{KKT_proof_2}
\hat{f}_l^i T_l^\prime(\hat{f}_l) + T_l(\hat{f}_l) \leq \hat{f}_{l+1}^i T_{l+1}^\prime(\hat{f}_{l+1}) + T_{l+1}(\hat{f}_{l+1}).
\end{equation}

From (\ref{KKT_proof_2}) it follows that $\hat{f}_l^i \leq \hat{f}_{l+1}^i$ $\forall i$,
for which $\hat{f}_l^i>0$, which
implies that $\hat{f}_l \leq \hat{f}_{l+1}$. It follows
that $T_l(\hat{f}_l) \leq T_l(\hat{f}_{l+1}) \leq T_{l+1}(\hat{f}_{l+1})$, which contradicts our assumption
that $T_l(\hat{f}_l)>T_{l+1}(\hat{f}_{l+1})$. Hence, $\forall l, 1 \leq l \leq L-1$, it holds that $T_l(\hat{f}_l) \leq T_{l+1}(\hat{f}_{l+1})$. 
}
\end{IEEEproof}

Furthermore, the following lemma establishes that there exists a `threshold link', in the following sense.  

\begin{lemma}
\label{homogeneous_2}
There exists a link $ M \in \cL$ for which: (i) $\forall l \leq M$,~$T_l(\hat{f}_l) \geq T_l(f_l^*)$  and (ii) $\forall n>M$,~$T_{n}(\hat{f}_{n}) \leq T_{n}(f_{n}^*)$.
\end{lemma}
\begin{IEEEproof}
\ifthenelse{\equal{\Context}{PAPER}}
{Assume by contradiction that there is no such link $M$.
We can split this into three (exhaustive) cases:
\begin{enumerate}
\item $T_l(f_l^*) \leq T_l(\hat{f}_l)~\forall l \in \cL$. Moreover, $\exists \bar{l}$, such that  $T_l(f_{\bar{l}}^*) < T_l(\hat{f}_{\bar{l}})$. 
\item $T_l(f_l^*) \geq T_l(\hat{f}_l)~\forall l \in \cL$. Moreover, $\exists \bar{l}$, such that  $T_l(f_{\bar{l}}^*) > T_l(\hat{f}_{\bar{l}})$. 
\item There are links $l,\bar{l} \in \cL, l < \bar{l}$, for which $T_l(\hat{f}_l) \leq T_l(f_l^*)$ and $T_{\bar{l}}(\hat{f}_{\bar{l}}) > T_{\bar{l}}(f_{\bar{l}}^*)$.
\end{enumerate}
In \cite{BO11}, we establish that each case leads to a contradiction.}
{
Assume by contradiction that there is no such link $M$.
We can split this into three (exhaustive) cases:
\begin{enumerate}
\item $T_l(f_l^*) \leq T_l(\hat{f}_l)~\forall l \in \cL$. Moreover, $\exists \bar{l}$, such that  $T_l(f_{\bar{l}}^*) < T_l(\hat{f}_{\bar{l}})$. 
\item $T_l(f_l^*) \geq T_l(\hat{f}_l)~\forall l \in \cL$. Moreover, $\exists \bar{l}$, such that  $T_l(f_{\bar{l}}^*) > T_l(\hat{f}_{\bar{l}})$. 
\item There are links $l,\bar{l} \in \cL, l < \bar{l}$, for which $T_l(\hat{f}_l) \leq T_l(f_l^*)$ and $T_{\bar{l}}(\hat{f}_{\bar{l}}) > T_{\bar{l}}(f_{\bar{l}}^*)$.
\end{enumerate}


Cases 1 and 2 respectively indicate that $\sum_{l \in \cL} f_l^* < \sum_{l \in \cL} \hat{f}_l$ and $\sum_{l \in \cL} f_l^* > \sum_{l \in \cL}  \hat{f}_l$, each of which is a contradiction. 


Consider Case 3. Due to monotonicity, it implies that $\hat{f}_l \leq f_l^*$ and $\hat{f}_{\bar{l}} > f_{\bar{l}}^*$. 

First we prove that $f_l^*>0$. Assume by way of contradiction that $f_l^*=0$. Then, since $f_l^*\geq \hat{f}_l$, it follows that $\hat{f}_l =0$. Moreover, it follows that $f_{\bar{l}}^* = 0$ and $\hat{f}_{\bar{l}} = 0$. Indeed, otherwise, the KKT optimality conditions would induce:
\begin{equation}
\label{KKT_opt_contra}
T_{\bar{l}}(f_{\bar{l}}^*) + f_{\bar{l}}^* T_{\bar{l}}^{\prime}(f_{\bar{l}}^*) \leq T_{l}(f_{l}^*) + f_{l}^* T_{l}^{\prime}(f_{l}^*) = T_{l}(0)
\end{equation}

which, by Lemma \ref{homogeneous_1}, is a contradiction. 
Similarly, the KKT condition (\ref{kut1}) would induce, for all $i$ for which $f_{\bar{l}}^i>0$:
\begin{equation}
\label{KKT_opt_contra_2}
T_{\bar{l}}(\hat{f}_{\bar{l}}) + \hat{f}_{\bar{l}}^i T_{\bar{l}}^{\prime}(\hat{f}_{\bar{l}}) \leq T_{l}(\hat{f}_{l}) + \hat{f}_{l}^i T_{l}^{\prime}(\hat{f}_{l}) = T_{l}(0)
\end{equation}

also which, by Lemma \ref{homogeneous_1}, is a contradiction. 
Hence, we conclude that $f_l^*>0$. 
Now suppose that $f_{\bar{l}}^*>0$. Then, since $f_l^*>0$, the KKT optimality conditions imply:



\begin{equation}
\label{KKT_proof_3}
f_l^* T_l^{\prime}(f_l^*) + T_l(f_l^*) = f_{\bar{l}}^* T_{\bar{l}}^{\prime}(f_{\bar{l}}^*) + T_{\bar{l}}(f_{\bar{l}}^*).  
\end{equation}

Hence, from (\ref{KKT_proof_3}) and Lemma \ref{homogeneous_1} we conclude that $f_l^* \geq f_{\bar{l}}^*$. For $f_{\bar{l}}^* = 0$, clearly $f_l^* \geq f_{\bar{l}}^*$. Thus we always have $f_l^* \geq f^*_{\bar{l}}$. 
Analogously, since $T_{\bar{l}}(\hat{f}_{\bar{l}}) > T_{\bar{l}}(f_{\bar{l}}^*)\geq 0$, the KKT condition (\ref{kut1}) for link $l,\bar{l}$ induces,
for all $i$ for which $\hat{f}_{\bar{l}}^i>0$:
\begin{equation}
\label{KKT_proof_3_Nash}
\hat{f}_l^i T_l^{\prime}(\hat{f}_l) + T_l(\hat{f}_l) \geq \hat{f}_{\bar{l}}^i T_{\bar{l}}^{\prime}(\hat{f}_{\bar{l}}) + T_{\bar{l}}(\hat{f}_{\bar{l}}),
\end{equation}

thus, from (\ref{KKT_proof_3_Nash}) it follows that $\hat{f}_l^i\geq \hat{f}_{\bar{l}}^i,~\forall i$ for which $\hat{f}_{\bar{l}}^i>0$. It
thus follows
that $\hat{f}_l \geq \hat{f}_{\bar{l}}$.
Therefore:

\begin{equation}
\label{relative}
0 \leq f_{\bar{l}}^* < \hat{f}_{\bar{l}} \leq \hat{f}_l \leq f_l^*.
\end{equation}

From (\ref{KKT_proof_3}) and (\ref{relative}) it follows that:

\begin{equation}
\hat{f}_l T_l^{\prime}(\hat{f}_l) + T_l(\hat{f}_l) 
< \hat{f}_{\bar{l}} T_{\bar{l}}^{\prime}(\hat{f}_{\bar{l}}) + T_{\bar{l}}(\hat{f}_{\bar{l}}).  
\end{equation}

Thus:

\begin{equation}
\label{contra_proof}
\hat{f}_l T_l^{\prime}(\hat{f}_l) - \hat{f}_{\bar{l}} T_{\bar{l}}^{\prime}(\hat{f}_{\bar{l}})  
< T_{\bar{l}}(\hat{f}_{\bar{l}}) -T_l(\hat{f}_l). 
\end{equation}



On the other hand, let $I_{\bar{l}} = \{ i \mid \hat{f}_{\bar{l}}^i>0 \} $. 
Taking the sum over all such $i$, it follows from (\ref{KKT_proof_3_Nash}) that:

\begin{equation}
\label{KKT_proof_6}
\sum_{i \in I_{\bar{l}}} \hat{f}_l^i T_l^{\prime}(\hat{f}_l) + \mid I_{\bar{l}} \mid \cdot T_l(\hat{f}_l) \geq \hat{f}_{\bar{l}} T_{\bar{l}}^{\prime}(\hat{f}_{\bar{l}}) +  \mid I_{\bar{l}} \mid \cdot T_{\bar{l}}(\hat{f}_{\bar{l}}).  
\end{equation}

Thus,
it follows that:

\begin{equation}
\label{KKT_proof_16}
\hat{f}_l T_l^{\prime}(\hat{f}_l) + \mid I_{\bar{l}} \mid \cdot T_l(\hat{f}_l) \geq \hat{f}_{\bar{l}} T_{\bar{l}}^{\prime}(\hat{f}_{\bar{l}}) +  \mid I_{\bar{l}} \mid \cdot T_{\bar{l}}(\hat{f}_{\bar{l}}).  
\end{equation}

It follows from (\ref{KKT_proof_16}) that: 

\begin{align}
\label{KKT_proof_7}
\hat{f}_l T_l^{\prime}(\hat{f}_{l}) - \hat{f}_{\bar{l}} T_{\bar{l}}^{\prime}(\hat{f}_{\bar{l}})
&\geq \mid I_{\bar{l}} \mid \cdot \left( T_{\bar{l}}(\hat{f}_{\bar{l}})  - T_l(\hat{f}_l) \right) \nonumber \\
&\geq 
 T_{\bar{l}}(\hat{f}_{\bar{l}}) - T_l(\hat{f}_l) 
\end{align}

which contradicts (\ref{contra_proof}). Hence, there do not exist links $l,\bar{l} \in \cL, l<\bar{l}$ for which $T_l(\hat{f}_l) \leq T_l(f_l^*)$ and $T_{\bar{l}}(\hat{f}_{\bar{l}}) > T_{\bar{l}}(f_{\bar{l}}^*)$. We conclude that there exists a link $M \in \cL $, as described in the lemma. 

} 
\end{IEEEproof}
\ifthenelse{\equal{\Context}{PAPER}}
{\subsection{Prices of Anarchy and Selfishness}
}
{\subsection{Prices of Anarchy, Selfishness and Isolation}
}
Define $\hat{\cJ}$ as the set of all cost vectors that correspond to a Nash equilibrium.
The {\em Price of Anarchy (PoA)} \cite{Koutsoupias99} is defined as
\begin{equation}
PoA = \sup_{\mathbf{J} \in \hat{\cJ}}\frac{J_{sys}}{J_{sys}^*}.
\end{equation}
Note that, in our routing game, there exists a unique NEP \cite{ORS93}, with social cost $\hat{J}_{sys}$. Hence, in our setting, the PoA is defined as
\begin{equation}
PoA = \frac{\hat{J}_{sys}}{J_{sys}^*}.
\end{equation}
For quantifying the degradation of performance under a {\em cooperative game} scenario, i.e., {\em solely} due to the selfish nature
of the decision makers, we introduce the following concept.
\begin{definition}
The {\em Price of Selfishness (PoS)} is the ratio between the {\em worst} expected social cost of any Nash Bargaining Solution $\mathbf{\tilde{g}}({\cG},\mathbf{\hat{J}})$, for which the disagreement point is a NEP, and the optimal social cost, i.e.,
\begin{equation}
PoS = \sup_{\substack{\mathbf{\tilde{g}}({\cG},\mathbf{\hat{J}}) \\ \mathbf{\hat{J}} \in \hat{\cJ}}} \frac{\mathbf{\tilde{g}}({\cG},\mathbf{\hat{J}})_{sys}}{J_{sys}^*}.
\end{equation}
Due to Theorem \ref{the:t1}, in our routing game, $\mathbf{\hat{J}}$, is unique. We denote the social cost of the unique NBS of the bargaining game $(\cG, \mathbf{\hat{J}})$, as $\tilde{J}_{sys}$. Hence, the Price of Selfishness in our game is defined as
\begin{equation}
PoS = \frac{\tilde{J}_{sys}}{J_{sys}^*}.
\end{equation}
\end{definition}
Due to the axiom of Individual Rationality ({\em N1}), the {\em PoS} is never higher than the {\em PoA} and potentially could be much lower.
The difference between the two is an issue of major importance, as it indicates how much gain (if at all) is accomplished by allowing
the decision makers to bargain and reach binding agreements.
\ifthenelse{\equal{\Context}{PAPER}}
{
}
{ 
For completeness, we also introduce a further concept, as follows:
\begin{definition}
The {\em Price of Isolation (PoI)} is the largest ratio of the social cost at a 
Nash Equilibrium and the expected social cost of its corresponding Nash Bargaining Scheme,
i.e.,
\begin{equation}
PoI = \sup_{\mathbf{\hat{J}} \in \hat{\cJ}} \frac{\hat{J}_{sys}}{\mathbf{\tilde{g}}({\cG},\mathbf{\hat{J}})_{sys}}.
\end{equation}
\end{definition}
In our routing game, 
\begin{equation}
PoI = \frac{\hat{J}_{sys}}{\tilde{J}_{sys}}.
\end{equation}
The {\em PoI} quantifies the degradation of performance paid due to the inability of the (selfish)
decision makers to communicate and reach an agreement. We note that, for a given game (e.g., the routing game considered in this study), worst-case
values of the {\em PoA}, {\em PoS} and {\em PoI} may be obtained in (three) different scenarios.
} 

\section{Homogeneous Costs}\label{sec:PoS_1}
It is well known that the Price of Anarchy can assume large values and in section \ref{sec:ident}, we present a
generic example within our framework, for which the {\em PoA} is unbounded. 
However, we proceed to establish that, 
such deficiency of performance 
can be mitigated
through bargaining. Specifically, we show that, with homogeneous costs, the Nash Bargaining Scheme 
strictly improves the system's performance, and strictly lowers the cost of each individual user, unless of course $PoA = 1$, in which case $PoS=PoA=1$. 
In order to establish that the Nash Bargaining Scheme strictly improves the performance of every user, we constructively design a strategy profile $\bf \bar{g}$, whose corresponding cost vector strictly satisfies Axiom {\em N1}, i.e., $\forall i,~\bar{g}^i<\hat{J}^i$.
This effectively proves that our $N$-player bargaining problem is {\em essential}. It then follows from Theorem \ref{the:t2} that the NBS strictly lowers the system's cost, compared to the system costs at the Nash equilibrium.

Our goal is to bring forth an initial feasible routing strategy profile, $\bf \bar{f}$, and constructively adjust it by exchanging flow between users, such that its corresponding cost vector will 
strictly satisfy Axiom {\em N1}. 
Initially, define $\bf \bar{f}$ as the routing strategy profile where all the users send their flow proportionally with regard to the system optimum, namely:
\begin{equation}
\label{proportionally_rout}
\forall_{i \in \cN, l \in \cL} ~\bar{f}_l^i = \frac{r^i}{R}f_l^*. 
\end{equation}

The corresponding users' costs thus equal\footnote{The process (algorithm), which will be described in 
Lemma \ref{alg_NBS}, consists of exchanging flows between users. After each such exchange, with a slight abuse of notation, we will denote the new cost of each user by $\bar{J}^i$ and avoid indexing the steps. $\bar{J}^i$, therefore, may change after each exchange of flow between users.}:
\begin{equation}
\label{proportionally}
\bar{J}^i(\bar{f}_l^i, \bar{f}_l) = \frac{r^i}{R}J^*_{sys} = \frac{r^i}{R} \cdot \sum_{l\in \cL} f_l^* T_l(f_l^*).
\end{equation}

From (\ref{proportionally_rout}) it is clear that the users send their aggregated flow such that it equals the system optimum, yet this does not imply that Axiom {\em N1} is strictly satisfied. Specifically, 
we need all the users to strictly lower their costs with respect to the NEP. However, when sending flow according to (\ref{proportionally_rout}),
it may happen that some users increase their cost in comparison to the NEP. Therefore,
we specify a process (algorithm), where users exchange flow between themselves, such that $\bf \bar{J}$
strictly satisfies Axiom {\em N1}, 
while maintaining system optimality. Note that
exchanges of flow between users shall not affect the aggregated flow on each of the links, hence the system 
optimum is still reached after each exchange.  

Define $\bar{G}^i \triangleq \hat{J}^i - \bar{J}^i$. 
We divide the users up into two sets. One set ($\bar{G}^+$) contains the users with a lower cost at $\bar{J}^i(\bar{f}_l^i,\bar{f}_l)$, compared to the NEP, i.e., $k \in G^+$ if $\bar{G}^k > 0$. The complementary set ($G^-$) contains the users with a higher or equal cost at $\bar{J}^i(\bar{f}_l^i,\bar{f}_l)$, in comparison to the NEP, i.e., $k \in G^-$ if $\bar{G}^k \leq 0$. 
From (\ref{proportionally}), the following holds:
\begin{align}
\label{proof_1}
&\sum_{i \in  G^+} \frac{r^i}{R} \cdot \sum_{l \in \cL} f_l^* T_l(f_l^*) + \sum_{j \in  G^-} \frac{r^j}{R} \cdot \sum_{l \in \cL} f_l^* T_l(f_l^*) 
\\
\leq
&\sum_{i \in  G^+} \sum_{l \in \cL} \hat{f}_l^i T_l(\hat{f}_l) + \sum_{j \in  G^-} \sum_{l \in \cL} \hat{f}_l^j T_l(\hat{f}_l) \nonumber ,
\end{align}
where the left side of (\ref{proof_1}) equals the optimal system cost and the right side equals the system cost at the NEP. 

From (\ref{proof_1}) we derive that:
\begin{align}
\label{proof_2}
&\sum_{i \in  G^+} \left[  \sum_{l \in \cL} \hat{f}_l^i T_l(\hat{f}_l) - \frac{r^i}{R} \cdot \sum_{l \in \cL} f_l^* T_l(f_l^*) \right] 
\\
\geq
&\sum_{j \in  G^-} \left[ \frac{r^j}{R} \cdot \sum_{l \in \cL} f_l^* T_l(f_l^*) -
\sum_{l \in \cL} \hat{f}_l^j T_l(\hat{f}_l) \right] \nonumber 
\end{align}
and from (\ref{proportionally}) and (\ref{proof_2}) it follows that:

\begin{equation}
\label{proof_3}
\sum_{i \in  G^+} \bar{G}^i \geq - \sum_{j \in  G^-} \bar{G}^j .
\end{equation}

Inequality (\ref{proof_3}) indicates that the overall gain from the users in $G^+$ exceeds the loss of the users in $G^-$. This implies that there can be an exchange of flow between users in $G^-$ and $G^+$, which translates into a exchange of cost, such that all users will end up in $G^+$, in which case Axiom {\em N1} is strictly satisfied.
However, in order for a user $m \in G^-$ to lower its cost and end up in $G^+$, another user $k \in G^+$ and a pair of links $l,n$ has to be found, such that $\bar{f}_l^k>0$, $\bar{f}_n^m>0$ and $T_l(f_l^*) < T_n(f_n^*)$. The following lemma proves that, as long as $G^-$ is not empty, such a user $k$ and links $l,n$ necessarily exist. 
\begin{lemma}
\label{switch}
In the game defined in Section~\ref{sec:model}, with homogeneous costs, consider instances for which the $PoA > 1$.
Assuming $G^-$ is nonempty, there always exists a tuple $(m,k,l,n)$ with users $m \in G^-$, $k \in G^+$ and links $l,n$, such that $\bar{f}_l^k>0$, $\bar{f}_n^m>0$ and $T_l(f_l^*) < T_n(f_n^*)$.  
\end{lemma}
\begin{IEEEproof}
Assume by way of contradiction that no such tuple can be found.
This implies that
$\forall m \in G^-, \forall k \in G^+$, $\forall l$ for which $\bar{f}_l^k>0$ and $\forall n$ for which $\bar{f}_n^m>0$, it holds that $T_l(f^*_l) \geq T_n(f^*_n)$\footnote{Since the $PoA>1$, from (\ref{proof_3}), it is straightforward that $G^+$ is non-empty.}. 
Thus, from Lemma \ref{homogeneous_1}, there exist two links $L^1, L^2$, $L^1 \leq L^2$, such that all users in $G^-$ send their flow on links $l=1, \ldots ,L^2$, all users in $G^+$ send their flow on links $l=L^1, \ldots ,L$, and for any two links $l,n$, $L^1 \leq l,n \leq L^2$ it holds that $T_l(f_l^*) = T_n(f^*_n)$.
Without loss of generality we can exchange flow between any users $ m \in G^-, k \in G^+$, on links $L^1\leq l \leq L^2$, such that after this exchange all users in $G^-$ send their flow on links $l=1, \ldots ,\bar{L}$, and all users in $G^+$ send their flow on links $l=\bar{L}, \ldots ,L$ for some link $L^1 \leq \bar{L} \leq L^2$. This exchange will not affect their cost.
Seeing that the aggregated flow of all the users continuously brings about the optimum, it follows that:
\begin{equation}
\label{total_plus}
\sum_{i \in G^-} r^i = \sum_{l =1}^{\bar{L}-1} f_l^* + f_{\bar{L}}^{-*}
~\mbox{and}~ \sum_{i \in G^+} r^i = f_{\bar{L}}^{+*} +  \sum_{l = \bar{L}+1}^{L} f_l^{*},
\end{equation}
where $f_{l}^{-*}$ and $f_{l}^{+*}$ respectively represent the amount of flow that users in $G^-$ and $G^+$ send on link $l \in \cL$ at the system optimum, i.e., $f^{+*}_{l} + f^{-*}_{l} = f^*_{l}$.
Furthermore, from Lemma \ref{homogeneous_2}, we know that there exists a threshold link, {\em M}, for which (i) $T_l(\hat{f}_l) \geq T_l(f_l^*),~\forall l \leq M$, and (ii) $T_n(\hat{f}_n) \leq T_n(f_n^*),~\forall n>M$.
This leaves us with two cases:

\begin{enumerate}
\item  $M < \bar{L}$
\item  $M \geq \bar{L}$.
\end{enumerate}

Consider Case 1, denote the demand of all users in $G^+$ as $R^+$, i.e, $R^+ \triangleq \sum_{i \in G^+} r^i$ and the demand of all users in $G^-$ as $R^-$.
Also, define $\hat{f}_l^+ \triangleq \sum_{i \in G^+}  \hat{f}_l^i$ and $R^+_1 \triangleq \sum_{l=1}^{\bar{L}} \hat{f}_l^+$. 
Finally denote $\bar{J}^+ \triangleq \sum_{i \in G^+} \bar{J}^i$ and $\hat{J}^+ \triangleq \sum_{i \in G^+} \hat{J}^i$.
It follows from (\ref{total_plus}) that:
\begin{align}
\label{r_plus}
R^+ &= \sum_{l = \bar{L}}^{L} f_l^{+*} = \sum_{l \in \cL} \hat{f}_l^+ =  \sum_{l=1}^{\bar{L}} \hat{f}_l^+ + \sum_{l=\bar{L}+1}^{L} \hat{f}_l^+,
\end{align}
hence:
\begin{align}
\label{R_+}
R^+_1 &= \sum_{l=1}^{\bar{L}} \hat{f}_l^+ = f_{\bar{L}}^{+*} +  \sum_{l = \bar{L}+1}^{L} f_l^{*} - \sum_{l=\bar{L}+1}^{L} \hat{f}_l^+
\\ \label{R_+_2}
&= \sum_{l=\bar{L}+1}^L \left[ f_l^* - \hat{f}_l^+ \right] + f_{\bar{L}}^{+*}.
\end{align}
Note that we consider Case 1, thus $f_l^* \geq \hat{f}_l \geq \hat{f}_l^+$ for $l > M$. 
It follows from Lemma \ref{homogeneous_1}, Lemma \ref{homogeneous_2} and (\ref{R_+}) that:
\begin{align}
\label{plus}
\hat{J}^+ &= \sum_{l=1}^{\bar{L}} \hat{f}^+_l T_l(\hat{f}_l) + \sum_{l=\bar{L}+1}^L \hat{f}^+_l T_l(\hat{f}_l) 
\\ \nonumber
&\leq  \sum_{l=1}^{\bar{L}} \hat{f}^+_l \cdot T_{\bar{L}}(\hat{f}_{\bar{L}}) + \sum_{l=\bar{L}+1}^L \hat{f}^+_l T_l(\hat{f}_l) 
\\ \nonumber
&= R_1^+ \cdot T_{\bar{L}}(\hat{f}_{\bar{L}}) + \sum_{l=\bar{L}+1}^L \hat{f}^+_l T_l(\hat{f}_l) 
\\ \nonumber
&\leq \sum_{l=\bar{L}+1}^L f_l^* T_l(\hat{f}_l) + f_{\bar{L}}^{+*} \cdot T_{\bar{L}}(\hat{f}_{\bar{L}}) 
\\ \nonumber
&\leq \sum_{l=\bar{L}+1}^L f_l^* T_l(f_l^*) + f_{\bar{L}}^{+*} \cdot T_{\bar{L}}(f^*_{\bar{L}})  = \bar{J}^+.
\end{align}
The first inequality follows from Lemma \ref{homogeneous_1}, the second follows from Lemma \ref{homogeneous_1} and (\ref{R_+_2}) and the last inequality follows from Lemma \ref{homogeneous_2}. 
If $\hat{J}^+  \leq \bar{J}^+$, it implies that $\exists k \in G^+$, for which $\bar{G}^k \leq 0$, which is a contradiction to
the definition of $G^+$. 
Thus, Case 1 is not possible.

Now consider Case 2.
Denote $\bar{J}^- \triangleq\sum_{i \in G^-} \bar{J}^i$, $\hat{J}^- \triangleq\sum_{i \in G^-} \hat{J}^i$ and
consider a new routing strategy profile $\bf h$, where for any $l \in \cL$, $h_l = \hat{f}_l$. Moreover, at $\bf h$, all users in $G^-$ send their flow on links $l=1, \ldots ,K$ and all users in $G^+$ send their flow on links $l=K, \ldots ,L$ for some $K \in \cL$. In other words, the aggregated link flows at $\bf h$ are equal to the aggregated link flows at the NEP, however at $\bf h$, the users in $G^-$ send all their demand on the links with the lowest cost per unit of flow. 
Since $M \geq \bar{L}$ it follows from Lemma \ref{homogeneous_2} that $K \leq \bar{L} \leq M$.
To prove the lemma for Case 2 it is sufficient to establish that  $\bar{J}^- < \hat{J}^-$, which leads to a contradiction.
We first prove that  $\sum_{i \in G^-} J^i(\mathbf{h}) \leq \hat{J}^-$, whereafter we establish that $\bar{J}^- \leq \sum_{i \in G^-} J^i(\mathbf{h})$ and $\bar{J}^- < \hat{J}^-$.

Since $\forall l,~h_l=\hat{f}_l$, we can transform $\bf h$ into the NEP by repeatedly switching flows between the users in $G^-$ and $G^+$ until all users in $G^-$ send their demand according to $\mathbf{\hat{f}}^-$ and since $\forall l,~h_l=\hat{f}_l$, all users in $G^+$ send their demand according to $\mathbf{\hat{f}}^+$.
Note that the aggregated link flow during this process stays constant, thus it is clear that at every stage,
users in $G^-$ will only increase their cost by sending more flow on lower links, while users in $G^+$ only decrease their cost by sending more flow on the higher links. 
Therefore, $\sum_{i \in G^-} J^i(\mathbf{h}) \leq \hat{J}^-$. 

We continue to establish that $\bar{J}^- \leq \sum_{i \in G^-} J^i(\mathbf{h})$. 
Consider a different network, links $\bar{\cal{L}}={1,\ldots, \bar{L}}$ and total demand $\bar{R} = \sum_{l=1}^{\bar{L}} f^*_l$. Note that $\bar{R} = R^- + [f^*_{\bar{L}} - f^{-*}_{\bar{L}}]$.
We consider two different routing strategy profiles in this new network. The first is its system optimum, which we denote by $\bf g^*$. The second is the strategy profile where an amount of $[f^*_{\bar{L}} - f^{-*}_{\bar{L}}]$ is sent on link $\bar{L}$ and the rest of the demand, $R^-$, is sent according to $\bf h$.
Thus, due to the optimality of $\bf g^*$:
\begin{equation} \label{optimality_ineq}
\sum_{l=1}^{\bar{L}} g^*_lT_l(g^*_l) 
\leq \sum_{l=1}^{K}h_lT_l(h_l) + [f^*_{\bar{L}} - f^{-*}_{\bar{L}}]T_{\bar{L}}(f_{\bar{L}}^*).
\end{equation}
However, from the KKT conditions (\ref{kut1}), it follows that for any link $l \in \bar{\cal{L}}$, $g^*_l = f^*_l$.
Hence, from (\ref{optimality_ineq}) we get that 
\begin{align} 
\bar{J}^- &= \sum_{l=1}^{\bar{L}-1}f^*_lT_l(f_l^*) + f^{-*}_{\bar{L}}T_{\bar{L}}(f_{\bar{L}}^*) 
\\ \nonumber
&= \sum_{l=1}^{\bar{L}} g^*_lT_l(g^*_l) - [f^*_{\bar{L}} - f^{-*}_{\bar{L}}]T_{\bar{L}}(f_{\bar{L}}^*) 
\\ \nonumber
&\leq \sum_{l=1}^{K}h_lT_l(h_l),
\end{align}
Thus, $\bar{J}^- \leq \sum_{i \in G^-} J^i(\mathbf{h})$ and consequently, $\bar{J}^- \leq \hat{J}^-$.
If $\bar{J}^- < \hat{J}^-$, then $\exists k \in G^-$, for which $\bar{G}^k > 0$, which is a contradiction to the definition of $G^-$. 

Now consider $\hat{J}^- = \bar{J}^-$. 
Since $\bar{L} \leq M$, it follows from (\ref{optimality_ineq}) and the definition of $\bf g$ and $\bf h$ that for any link $l \leq \bar{L}$, $T_l(\hat{f}_l) = T_l(f^*_l)$. Moreover, $K=\bar{L}$.
Consequently, according to Case 2 and 3 in Lemma \ref{homogeneous_2}, it follows that $\forall l,n \in \cL$, $T_l(\hat{f}_l) = T_l(f^*_l)$ and, the Price of Anarchy equals 1, which is a contradiction to the conditions of the lemma.
Thus, Case 2 is not possible either. We therefore conclude that there always exists a tuple $(m,k,l,n)$ such as described in the lemma, unless $G^-$ is empty.  
\end{IEEEproof}


In the following lemma, we establish a process (which, in fact, is a computationally efficient
algorithm), which finds tuples and exchanges flow between users such that all users will end up in $G^+$.
%
%
%
%
%
\begin{lemma}
\label{alg_NBS}
In the game defined in Section~\ref{sec:model}, with homogeneous costs, consider instances for which the $PoA > 1$.
There exists a bargained strategy profile, which equals the system optimum and strictly satisfies Axiom {\em N1}. 
\end{lemma}
\begin{IEEEproof}
Lemma \ref{switch} shows that we are always able to find a tuple ($m,k,l,n$), unless $G^-$ is empty and all users are in $G^+$, in which case Axiom {\em N1} is strictly satisfied. After finding such a tuple, user $m \in G^-$ exchanges flow with another user $k \in G^+$, thereby lowering its cost until one of the following events occurs: 

\begin{enumerate}
\item $\bar{f}_l^k = 0$.
\item $\bar{f}_n^m = 0$.
\item $\bar{G}^k = \epsilon$, for some small enough $\epsilon>0$.
\item $m \in G^+$. 
\end{enumerate}
We choose $\epsilon$ as a threshold, whereafter $k$ refrains from exchanging its flow. 

Unless $G^-$ is empty, each time one of the events {\em 1-4} occurs, a new tuple can be found as explained in Lemma \ref{switch}.
Furthermore, from the strict inequality of (\ref{proof_3}) it follows that there is enough cost to be transfered from users in $G^+$ to users in $G^-$, such that, for a small enough $\epsilon$, all users will end up in $G^+$. 
We therefore establish the following algorithm that increments over the links $l \in \cL$ and finds tuples $(m,k,l,n)$ until $G^-$ is emptied.
 
Denote $l^+$ as the link for which $\exists k \in G^+$ such that $\bar{f}_{l^+}^k>0$ and $\forall l<l^+, \forall i \in G^+$ it holds that $\bar{f}_{l}^i = 0$. Likewise, define $l^-$ as the link for which $\exists m \in G^-$ such that $\bar{f}_{l^-}^m>0$ and $\forall l>l^-, \forall j \in G^-$ it holds that $\bar{f}_{l}^j = 0$. 

In the first step, set $l^+ = 1$ and $l^- = L$. 
When two users $m,k$ are found, such as described in Lemma \ref{switch}, they exchange flow till one of the events {\em 1-4} occurs. Afterwards, a new tuple $(m',k',l^+,l^-)$ is found for users $m',k'$. 

When no user $k \in G^+$ can be found on link $l^+$, we increment the link number such that $l^+$ $\Leftarrow$ $l^+ + 1$. The algorithm now looks for a tuple $(m,k,l^+ + 1,l^-)$.  
Similarly, when no users $m \in G^-$ can be found on link $l^-$, we decrement the link number, such that, $l^-$ $\Leftarrow$ $l^- - 1$. The algorithm then looks for a tuple $(m,k,l^+,l^- - 1)$. 
Once $l^+ = l^-$, no such tuple can be found, hence, by Lemma \ref{switch}, $G^-$ is empty. Furthermore, the algorithm completes within a final number of steps (as formally established by
Proposition \ref{complexity_NBS} that follows). 

After the algorithm, the set $G^-$ is empty and we obtain a cost vector that strictly satisfies Axiom {\em N1}. 
Furthermore, during every step of the algorithm the aggregated flow on the links is equal to the flow at the system optimum.
\end{IEEEproof}
We are now able to state the following theorem. 
\begin{theorem}
\label{thm:essential}
In the game defined in Section~\ref{sec:model}, with homogeneous costs, consider instances for which the Price of Anarchy is strictly larger than 1, i.e., $PoA > 1$.
\begin{enumerate}
\item The corresponding $N$-player bargaining problem is essential.
\item At the outcome of the NBS, each user strictly decreases its cost. Thus, Price of Selfishness is strictly smaller than the Price of Anarchy ($PoS < PoA$).
\end{enumerate} 
\end{theorem}
\begin{IEEEproof}
In Lemma \ref{alg_NBS} we established a bargained strategy profile, $\bf \bar{f}$, whose cost vector is socially optimal and strictly satisfies Axioms {\em N1}. This completes the first claim of the theorem.
Since the corresponding $N$-player bargaining problem is {\em essential}, 
$\prod_{i=1}^N \left( \hat{J}^i - J^i(\mathbf{\bar{f}}) \right) >0,$
and as a result of Theorem \ref{the:t2} it follows that for the Nash Bargaining Solution, $\bf \tilde{g}$, 
$$\prod_{i=1}^N \left( \hat{J}^i - \tilde{g}^i \right) \geq \prod_{i=1}^N \left( \hat{J}^i - J^i(\mathbf{\bar{f}}) \right) >0.$$
Thus, at the NBS each user strictly decreases its cost and from (\ref{social_cost}), it follows that $PoS < PoA$, hence establishing the second claim of the theorem.
\end{IEEEproof}

Note that, since this cost vector is socially optimal, it also satisfies Axiom {\em N2} (Pareto Optimality). Thus, the scheme described in the proof of Lemma \ref{alg_NBS} effectively constitutes an efficient algorithm for computing a cost vector that satisfies Axiom {\em N1} and Axiom {\em N2}. 
\begin{proposition}
\label{complexity_NBS}
Given the NEP, the process described in Lemma \ref{alg_NBS} is an $O(N \cdot L)$ algorithm for computing a feasible cost vector which satisfies Axioms {\em N1} and {\em N2}.
\end{proposition}
\begin{IEEEproof}
\ifthenelse{\equal{\Context}{PAPER}}
{See \cite{BO11}.}
{
The ordering of the links at the optimum takes $O(L)$ time. The events {\em 1-4}, as described in Lemma \ref{alg_NBS}, may occur $O(N)$ times for each pair of links, hence the time it takes for each pair of links, $(l^+,l^-)$, to find multiple tuples of users, is $O(N)$. The pointers $l^+$ and $l^-$ increment over the links, therefore tuples have to be found for a maximum of $L$ links. Since, each exchange of flow between users takes $O(1)$ time, the total computing time is $O(N \cdot L)$. 
} 
\end{IEEEproof}
According to Theorem \ref{thm:essential} it follows that all users (and the system), stand to gain from bargaining.
However, it remains an open question as to how much the cost of the system is reduced.
In the following sections we will show that, for certain instances, through bargaining, the users bring the system to optimality, thereby completely overcoming the deficiency implied by their selfish behavior. 
Specifically, we shall establish that in a 2-user system as well as
in an $N$-user system where all users have equal demands, 
the NBS brings about the social optimum, i.e., $PoS = 1$.

\subsection{Two Users}
\label{sec:two_users}
In this section we consider a system that consists of two users, i.e., $N=2$. We denote the two users in the system as $i$ and $j$.
We show that, with homogeneous costs, the Nash Bargaining Scheme always (i.e., with probability 1) brings the system to its social optimum. Moreover, we will establish that this is done through a single bargained strategy profile, chosen with probability 1. 
Thus, through bargaining, the deficiency of the network at the NEP can be overcome entirely.
In order to achieve this, we constructively design a bargained strategy profile, whose corresponding cost vector is socially optimal and, at the same time, complies with Axioms {\em N1-N5}. Thus, it follows from Theorem \ref{the:t2} that this cost vector is the unique solution of the NBS. We then choose this bargained strategy profile with probability 1, hence the {\em PoS} is always equal to 1. 

As a result of Lemma \ref{alg_NBS}, there might exist a range of system-optimal cost vectors for which Axioms {\em N1} and {\em N2} are satisfied. However, we focus on a particular cost vector that is system optimal and also complies with Axioms {\em N3-N5}. To do this, we first exhaustively describe the set of all cost vectors that satisfy Axiom {\em N1}. Denote this set as $\cG_{N1} \subseteq \cG$. We then focus on a specific cost vector within this set.
The following lemma describes two instances of system-optimal cost vectors for which Axiom {\em N1} is only weakly (i.e., not strictly) satisfied. Hence, these cost vectors lie on the boundary of $\cG_{N1}$.
\begin{lemma}
\label{lem:opt_help}
For any user $k$, there exists a system-optimal routing strategy profile, $\bf \bar{f}$, for which $\bar{J}^k \geq \hat{J}^k$. 
\end{lemma}
\begin{IEEEproof}
We consider user $i$, since the proof is symmetric for user $j$. 
At the NEP, we split the links into two sets: $\cL^+ = \{l \in \cL | \hat{f}_l^j \geq f_l^*\}$ and $\cL^- = \{l \in \cL| \hat{f}_l^j < f_l^*\}$. Now, consider a new routing strategy for user $i$, $\bf \bar{f}^i$ in which it ``fills'' up the links in $\cL^-$ according to the system optimum, starting from link $L$ upwards. 
After the filling process 
user $i$ reaches a link, $K \in \cL^-$ for which $\bar{f}^i_{K} \leq f^*_K - \hat{f}^j_K$ and  for any link $l>K,~l \in \cL^-, \bar{f}^i_{l} = f^*_l - \hat{f}^{j}_l$.
At the new routing strategy, $\bf \bar{f}^i$, $i$'s cost is equal to
\begin{equation}
\label{opt_nash_1}
J^i(\mathbf{\bar{f}^i, \hat{f}^j}) = \sum_{\substack{l > K \\ l \in \cL^-}} \left[f_l^*-\hat{f}_l^j \right]T_l(f_l^*) + \bar{f}_{K}^{i}T_{K}(\bar{f}_K^i+\hat{f}^{j}_{K}) 
\geq \hat{J}^i.
\end{equation}
The inequality follows from (\ref{eq:e31}).
We now change the routing strategy of user $j$ and construct a flow $\bf (\bar{f}^i,\bar{f}^j)$ that is system optimal and for which $J^i(\mathbf{\bar{f}^i,\bar{f}^j}) \geq J^i(\mathbf{\bar{f}^i, \hat{f}^j})$. By doing so we have constructed a feasible optimal routing profile $\bf \bar{f}$ for which $J^i(\mathbf{\bar{f}}) \geq \hat{J}^i$, hence proving the lemma.

We define the strategy $\bf \bar{f}^j$ as follows. On any link $l \in \cL$, $j$ sends an amount $\bar{f}^j_l$ such that $\bar{f}^j_l + \bar{f}^i_l = f^*_l$. Since for any link $l$, $\bar{f}_l^i \leq f_l^*$, this new routing strategy is feasible.
On any link $l \in \cL^-,~l > K$, $j$ does not increase its flow, i.e., $\bar{f}^j_l = \hat{f}^j_l$. Furthermore, on link $K$, $\bar{f}^j_K \geq \hat{f}^j_K$.
As a result,  
\begin{align}
\label{opt_nash_2}
J^i(\mathbf{\bar{f}^i, \bar{f}^j}) &= \sum_{\substack{l > K \\ l \in \cL^-}} \left[f_l^*-\bar{f}_l^j \right]T_l(f_l^*) + \bar{f}_{K}^{i}T_{K}(\bar{f}_K^i+\bar{f}^{j}_{K}) 
\\ \nonumber
&\geq \sum_{\substack{l > K \\ l \in \cL^-}} \left[f_l^*-\hat{f}_l^j \right]T_l(f_l^*) + \bar{f}_{K}^{i}T_{K}(\bar{f}_K^i+\hat{f}^{j}_{K}).
\end{align}
Hence, from (\ref{opt_nash_1}) and (\ref{opt_nash_2}), $J^i(\mathbf{\bar{f}^i, \bar{f}^j}) \geq J^i(\mathbf{\bar{f}^i, \hat{f}^j}) \geq \hat{J}^i$.
\end{IEEEproof}
With Lemma \ref{lem:opt_help}, we can now exhaustively describe $\cG_{N1}$.
\begin{lemma}
\label{lem:triangle}
The set of cost vectors that satisfy Axiom {\em N1} is equal to a triangle with vertices at 
$(\hat{J}^i, J_{sys}^* - \hat{J}^i)$, $(J_{sys}^* - \hat{J}^j, \hat{J}^j)$ and $(\hat{J}^i, \hat{J}^j)$. (See Figure \ref{fig:NBS}).
\end{lemma}
\begin{IEEEproof}
\begin{figure}[h!]
  \centering
    \includegraphics[width=0.35\textwidth]{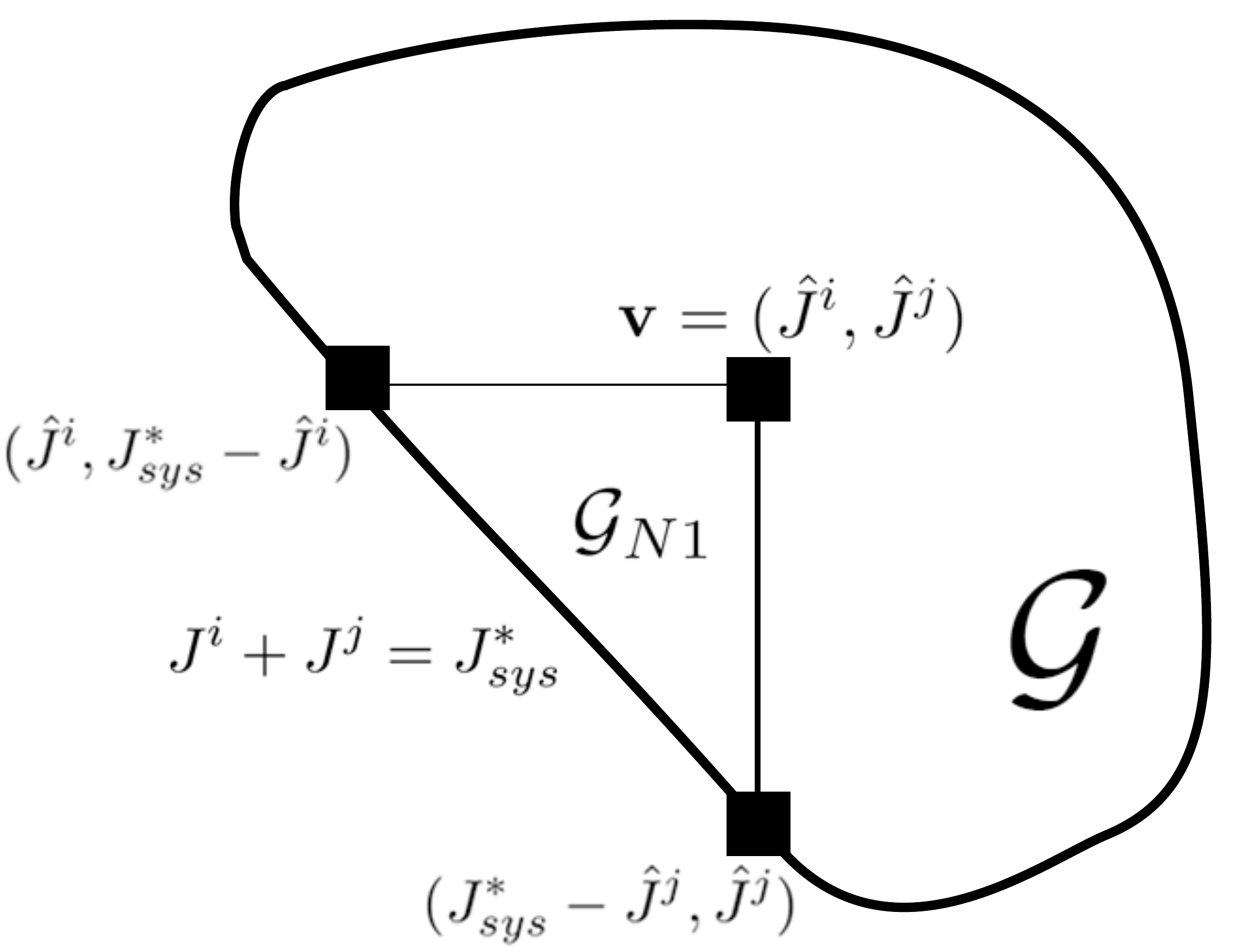}
  \caption{Set of bargainable costs.}
\label{fig:NBS}
\end{figure}
From Lemma \ref{lem:opt_help}, it is immediate that the cost vectors $\mathbf{x} = (\hat{J}^i, J_{sys}^* - \hat{J}^i)$, $\mathbf{y} = (J_{sys}^* - \hat{J}^j, \hat{J}^j)$ and $\mathbf{z} = (\hat{J}^i, \hat{J}^j)$ all lie in $\cG_{N1}$. 
Moreover, since $\cG$ is convex, the three edges $(\mathbf{x},\mathbf{y})$, $(\mathbf{x},\mathbf{z})$ and $(\mathbf{y},\mathbf{z})$ also lie in $\cG_{N1}$.
For any cost vector $\bf g$ that lies on $(\mathbf{x},\mathbf{z})$ or $(\mathbf{y},\mathbf{z})$, Axiom {\em N1} is weakly satisfied. Moreover, since $g^i + g^j \geq J_{sys}^*$ for any $\mathbf{g} \in \cG$, the 
edges $(\mathbf{x},\mathbf{z})$ and $(\mathbf{y},\mathbf{z})$ describe the boundaries of $\cG_{N1}$. 
Finally, the edge $(\mathbf{x},\mathbf{y})$ is equal to the line $g^i + g^j = J^*_{sys}$ where $g^i \leq \hat{J}^i$ and $g^j \leq \hat{J}^j$.
It is straightforward that any cost vector outside the edges of the triangle either does not satisfy Axiom {\em N1} or is not feasible.
\end{IEEEproof}
After describing $\cG_{N1}$, we now make use of the remaining Axioms {\em N2-N5} to identify a specific cost vector in $\cG_{N1}$ which is equal to the unique NBS.
\begin{lemma}
\label{lem:NBS_equiv}
Consider the bargaining problem $(\cG,\mathbf{\hat{J}})$. The Nash Bargaining Solution, 
$\mathbf{\tilde{g}}(\cG,\mathbf{\hat{J}})$,
corresponds to the feasible, system-optimal cost vector for which 
$$\hat{J}^i -\tilde{g}^i(\cG,\mathbf{\hat{J}}) = \hat{J}^j -\tilde{g}^j(\cG,\mathbf{\hat{J}}).$$
\end{lemma}
\begin{IEEEproof}
From Axiom {\em N1}, we know that $\mathbf{\tilde{g}}(\cG,\mathbf{\hat{J}}) \in \cG_{N1}$.
Since $\cG_{N1} \subseteq \cG$, from Axiom {\em N5}, it follows that
\begin{equation}
\label{NBS_equal}
\mathbf{\tilde{g}}(\cG,\mathbf{\hat{J}}) = \mathbf{\tilde{g}}(\cG_{N1},\mathbf{\hat{J}}).
\end{equation}
Thus, in order to prove the lemma, it suffices to look at the bargaining problem $(\cG_{N1},\mathbf{\hat{J}})$. 
Now consider the linear transformation of $\cG_{N1}$ such that 
$$\bar{\cG}_{N1} = \{g^i - \hat{J}^i, g^j - \hat{J}^j|(g^i,g^j) \in \cG_{N1}\}.$$
After the transformation, the vectors $\mathbf{x}$, $\mathbf{y}$, $\mathbf{\hat{J}}$ from Lemma \ref{lem:triangle}, correspond to respectively, $(0, J_{sys}^* - \hat{J}_{sys})$, $(J_{sys}^* - \hat{J}_{sys}, 0)$ and $(0,0)$. 
Thus, the bargaining problem $(\bar{\cG}_{N1},(0,0))$ is symmetric for both users and according to Axiom {\em N3}, $\tilde{g}^i(\bar{\cG}_{N1},(0,0)) = \tilde{g}^j(\bar{\cG}_{N1},(0,0))$. 
Moreover, from Axiom {\em N2}, $\mathbf{\tilde{g}}(\bar{\cG}_{N1},(0,0))$ lies on the Pareto frontier of $\bar{\cG}_{N1}$, i.e., the edge $(\mathbf{x} - \mathbf{\hat{J}},\mathbf{y} - \mathbf{\hat{J}})$, hence 
 $\tilde{g}^i(\bar{\cG}_{N1},(0,0)) + \tilde{g}^j(\bar{\cG}_{N1},(0,0)) = J_{sys}^* - \hat{J}_{sys}$.
Consequently, 
$$\tilde{g}^i(\bar{\cG}_{N1},(0,0)) = \frac{J^*_{sys} - \hat{J}_{sys}}{2},~\tilde{g}^j(\bar{\cG}_{N1},(0,0)) = \frac{J^*_{sys} - \hat{J}_{sys}}{2}.$$
Due to Axiom {\em N4}, 
\begin{equation*}
\label{NBS_two}
\mathbf{\tilde{g}}(\cG_{N1},\mathbf{\hat{J}}) = (\tilde{g}^i(\bar{\cG}_{N1},(0,0)) +\hat{J}^i,~\tilde{g}^j(\bar{\cG}_{N1},(0,0)) +\hat{J}^j),
\end{equation*}
hence,
\begin{equation}
\label{NBS_solved}
\tilde{g}^i(\cG_{N1},\mathbf{\hat{J}}) = \frac{J^*_{sys} + \hat{J}^i - \hat{J}^j}{2},~\tilde{g}^j(\cG_{N1},\mathbf{\hat{J}}) = \frac{J^*_{sys} + \hat{J}^j - \hat{J}^i}{2}.
\end{equation}
Finally, from (\ref{NBS_equal}) and (\ref{NBS_solved}) it follows that
$\hat{J}^i - \tilde{g}^i(\cG,\mathbf{\hat{J}}) = \hat{J}^j - \tilde{g}^j(\cG,\mathbf{\hat{J}})$ 
and $\tilde{g}^i(\cG,\mathbf{\hat{J}}) + \tilde{g}^j(\cG,\mathbf{\hat{J}}) = J^*_{sys}$.
\end{IEEEproof}
As a result of Lemma \ref{lem:NBS_equiv}, we are able to prove the following theorem.
\begin{theorem}
\label{thm:two_users}
In the game defined in Section~\ref{sec:model}, with homogeneous costs and $N=2$, the Price of Selfishness equals $1$. Moreover, the outcome of the NBS is always (i.e., with probability $1$) socially optimal. 
\end{theorem}
\begin{IEEEproof}
As a result of Lemma \ref{lem:NBS_equiv}, we have established a bargained strategy profile, 
whose cost vector is socially optimal and complies with Axioms {\em N1-N5}.\footnote{Moreover, it is equal to both the {\em egalitarian solution} and {\em utilitarian solution} (see \cite{Myerson}).}
Henceforth, we can choose the constructed bargained strategy profile, with probability $1$. Therefore, the Price of Selfishness is equal to $1$ and the unique solution of the Nash Bargaining Scheme is always (i.e., with probability $1$) socially optimal.
\end{IEEEproof}
According to Theorem \ref{thm:two_users}, it is certainly worthwhile for both players, and for the entire system, to send their demands according to the NBS. 
Nevertheless, this result does not automatically extend to a system with more than two players. 
From Theorem \ref{thm:essential}, we know that all $N$-players strictly benefit by sending their demand according to the NBS, but system optimality is not guaranteed. 
Indeed, the following example brings a case in which $N>2$ and the NBS is not socially optimal. 
\begin{example}
Consider a network of three users and two parallel links. The demands of the users are $r^1 = 0.1, r^2 = r^3 = 7.45$ and the costs of the users equal
\begin{equation}\label{eq:NBS_costs}
J^i = \frac{f^i_1}{20 - f_1} + \frac{f^i_2}{10 - f_2},
\end{equation}
for $i=1,2,3$. 
According to Theorem \ref{the:t2}, we can find the NBS, by maximizing $(\hat{J}^1 - g^1)(\hat{J}^2 - g^2)(\hat{J}^3 - g^3)$ for any $\mathbf{g} \in \cG$. 
%
%
%
However, it follows that $\tilde{f}_1  = 11.17 < f^*_1$. Hence, the NBS does not bring the system to its optimum.
\end{example}
From the above example together with Theorem \ref{thm:essential} we conclude that for a general case of $N$-players, where $N>2$, it holds that $1 < PoS < PoA.$
It remains an open problem to tighten the bounds of the {\em PoS}.
Nevertheless, there exists other cases in which, for the general case of $N$-players, the NBS is system optimal, as will be shown in the next section.
\subsection{Identical Users}
\label{sec:ident}
\ifthenelse{\equal{\Context}{PAPER}}
{In \cite{BO11}, we provide a generic example of a two-link network, where all users have equal demands and the users' costs hold by Assumptions ({\em H1-H4}). In the example, we show that the {\em PoA} is unbounded by splitting the aggregate traffic demand among an increasing number of users. 
}
{
We will provide a generic example of a two-link network, for which all users have equal demands and the users' costs satisfy Assumptions {\em H1-H4}.  In the example, the {\em PoA} 
can easily be made larger by splitting the same aggregated traffic demand among an increasing number of users. 
\begin{example}
\label{ex:high_PoA}
Consider a set of $N$ players that route their traffic demands over two parallel links. Assume that their aggregated traffic demand 
is $1$ and that the (homogeneous) cost function of each user $i$ is as follows:
\begin{equation}\label{cost_example}
J^i = f^i_1 \cdot(1 + \epsilon f_1) + f^i_2 \cdot (f_2)^N, ~\forall i \in \cN,
\end{equation}
where $\epsilon>0$.
In \cite{ORS93} it is proven that, in such a case of ``symmetric users'', the flows at the NEP are such that 
$\hat{f}^i_l = \frac{\hat{f}_l}{N}, ~\forall i \in \cN, \forall l \in \cL$. 
Thus, from the KKT conditions (\ref{kut1}), if $f_1>0$ and $f_2>0$, it follows that
$$1 + \epsilon \hat{f}_1 \left(1 + \frac{1}{N} \right) = \left( 1+\frac{N}{N} \right) \cdot (\hat{f}_2)^N = 2(\hat{f}_2)^N .$$
Consequently, $(\hat{f}_2)^{N} > \frac{1}{2}$ and the system cost at the NEP is 
$$\hat{J}_{sys} \geq (f_2)^{N+1}>  \frac{1}{2}^{(N+1)/N} \geq \frac{1}{4}.$$
Moreover, if either $f_1 = 0$ or $f_2 = 0$, it follows from (\ref{cost_example}) and (\ref{social_cost}) that $\hat{J}_{sys} \geq 1$. Thus, $\hat{J}_{sys} \geq \frac{1}{4}$.

Now consider a new routing strategy profile $\bf \bar{f}$, where $\bar{f}_2 = (N+1)^{-1/N}$ and $\bar{f}_1 = 1 - (N+1)^{-1/N}$.
Thus,
\begin{align}
J^*_{sys} &\leq J_{sys}(\mathbf{\bar{f}}) = \bar{f}_1 \cdot(1 + \epsilon \bar{f}_1) + \bar{f}_2 \cdot (\bar{f}_2)^N,
\\ \nonumber
&= \left[1-(N+1)^{-\frac{1}{N}}\right] + \epsilon \left[1-(N+1)^{-\frac{1}{N}}\right]^2 +(N+1)^{-\frac{N+1}{N}}
\\ \nonumber
&< (1+\epsilon)\left[1-(N+1)^{-\frac{1}{N}}\right] +(N+1)^{-\frac{N+1}{N}} 
\\ \nonumber
&= 1+\epsilon -(N+1)^{-\frac{1}{N}} \left[\frac{\epsilon + N(1+\epsilon)}{N+1} \right]
\\ \nonumber
&< (1+\epsilon) \left[ 1- N(N+1)^{-\frac{N+1}{N}} \right],
\end{align}
which tends to 0 when $N\rightarrow \infty$. 
Thus, 
$$PoA \geq \frac{1/4}{(1+\epsilon) \left[ 1- N(N+1)^{-(N+1)/N} \right]},$$
which tends to $\infty$ when $N\rightarrow \infty$.
\end{example}
} 
In contrast to the unbounded {\em PoA}, the {\em PoS} will always be equal to 1 in these settings, 
as will be established in the next theorem. 
\begin{theorem}
\label{identical}
Consider the game defined in Section~\ref{sec:model}, where $r^i \equiv \frac{R}{N},~\forall i \in \cN$, and the users' costs satisfy Assumptions {\em H1-H4}. For the corresponding $N$-player bargaining problem, it holds that: 
\begin{enumerate}
\item Any solution $\bf \tilde{g}$ that satisfies Axioms {\em N2} and {\em N3}, is unique. 
\item The Price of Selfishness is always (i.e., with probability $1$) equal to 1.
\end{enumerate}
\end{theorem} 
\begin{IEEEproof}
Assume by contradiction that there exist two distinct bargained cost vectors, $\bf g,~h$, which both satisfy Axioms {\em N2-N3}.
Since all users have equal demand, it follows from \cite{ORS93} that, all users receive equal costs at the NEP.
Axiom {\em N3} states that all users should then receive equal costs at the NBS.
Thus, for any two players $i,j \in \cN$, $g^i = g^j$ and $h^i = h^j$.
Assume w.l.o.g. that $\sum_{i \in \cN} g^i > \sum_{i \in \cN} h^i$.
Consequently, 
$$g^i > h^i,~\forall i \in \cN.$$
However, by switching from $\bf g$ to $\bf h$, all users decrease their expected cost, which is in contradiction to Axiom {\em N2}. Thus, any bargained cost vector that satisfies Axioms {\em N2-3}, is unique.
This establishes the first part of the lemma.

In the same way as (\ref{proportionally_rout}), we define a bargained strategy profile $\bf \bar{f}$, where users send their flow proportionally to the social optimum of the system:
$\forall_{i \in \cN, l \in \cL} ~\bar{f}_l^i = \frac{r^i}{R}f_l^*$.

Similar to (\ref{proportionally}), the corresponding users' costs are: \\ 
\begin{equation}
\label{identical_cost_optimum}
\bar{J}^i(\bar{f}_l^i,\bar{f}_l) = \frac{r^i}{R} \bar{J}(\bar{f}_l^i,\bar{f}_l) = \frac{r^i}{R} \sum_{l \in \cL} f_l^{*} T_l(f_l^*) = \frac{J^*_{sys}}{N}.
\end{equation}
By combining the first part of the theorem together with Theorem \ref{thm:essential}, it follows that, any cost vector that satisfies Axioms {\em N2-3} is equal to the unique NBS. 
Thus, in order to demonstrate that the constructed bargained strategy profile also brings about the NBS, it remains to verify that it satisfies Axioms {\em N2} and {\em N3}.
\begin{itemize}
\item [N2] From (\ref{identical_cost_optimum}), it follows that the aggregated flow on the links brings about the optimum. 
By definition, the system optimum is Pareto optimal.
\item [N3] From (\ref{identical_cost_optimum}) it is clear that all users receive equal costs at the NBS, therefore Axiom {\em N3} is satisfied. 
\end{itemize}
Thus, for the case where the users' costs abide by assumptions {\em H1-H4}, we have constructed a cost vector $\bf \bar{J}$, which satisfies Axioms {\em N1-N5} and is equal to the system optimum. 
We can now choose the proposed bargained strategy profile with probability $1$, therefore the {\em PoS} is always (with probability $1$) equal to 1. 
\end{IEEEproof}
Note that, Theorem \ref{identical} relates to any bargaining scheme that satisfies Axioms {\em N2} and {\em N3}. Aside from the Nash Bargaining Solution, this includes many other solutions, such as the {\em egalitarian solution} or when $N=2$, the Kalai-Smorodinsky solution (see \cite{Myerson}). 

\subsection{Weighted Social Cost}\label{sec:soc_cost}
As mentioned, our focus lies on a social cost that is the sum of the users' costs, which is the common practice in the literature. Yet, it is of interest
to examine the sensitivity of the {\em PoS} to the choice of the social cost. To that end, we turn to consider a case where, while the costs of the
users are homogeneous,
the social cost assumes a more general
structure.  

Specifically, suppose that, from a system point of view, the performance of the users should not
be treated equally. Namely, for each user $i$, there is a coefficient (weight) $\alpha^i>0$ that captures the relative importance of its
performance. As these are relative weights, we set $\sum_{i \in \cN} \alpha^i =1$. The social cost is then: $\Jsys = \sum_{i \in \cN} \alpha^i \cdot \sum_l f^i_l T_l(f_l)$.
Note that 
$\Jsys$ stays convex.
We term the above as a {\em weighted social cost} and denote the Price of Selfishness of the non-weighted social cost as $\overline{PoS}$.
\begin{proposition}
\label{PoS_upper_bounded}
In the game defined in Section~\ref{sec:model}, with homogeneous costs and with a weighted social cost,
the {\em Price of Selfishness} is bounded by 
$$\frac{\min_{i \in \cN} \alpha^i}{\max_{i \in \cN} \alpha^i} \cdot \overline{PoS} \leq PoS \leq \frac{\max_{i \in \cN} \alpha^i}{\min_{i \in \cN} \alpha^i} \cdot \overline{PoS}.$$ 
\end{proposition}
\begin{IEEEproof}
\ifthenelse{\equal{\Context}{PAPER}}{
See \cite{BO11}.}
{
For the social weighted cost it holds that 
\begin{align}
\label{weighted_social_1}
\Jsys &= \sum_{i \in \cN} \alpha^i \cdot \sum_{l \in \cL} f^i_l T_l(f_l)
\\ \nonumber
&= \sum_{l \in \cL}  T_l(f_l) \sum_{i \in \cN} \alpha^i f^i_l
\\ \nonumber
&\leq \sum_{l \in \cL}  T_l(f_l) \max_{i \in \cN}\{\alpha^i\} \sum_{i \in \cN}  f^i_l
= \max_{i \in \cN}\alpha^i \sum_{l \in \cL}  f_l T_l(f_l). 
\end{align}
Similarly, by switching {\em max} with {\em min}, we get from (\ref{weighted_social_1}) that 
$\Jsys \geq \min_{i \in \cN}\alpha^i \sum_{l \in \cL}  f_l T_l(f_l)$.

By definition, the Price of Selfishness of the weighted case equals,
\begin{align*} 
PoS = \frac{\sum_{i \in \cN} \alpha^i \cdot \sum_l \tilde{f}^i_l T_l(\tilde{f}_l)}{\sum_{i \in \cN} \alpha^i \cdot \sum_l f^*_l T_l(f_l^*)}
&\leq 
\frac{\max_{i \in \cN} \alpha^i \cdot \sum_l \tilde{f}_l T_l(\tilde{f}_l)}{\min_{i \in \cN} \alpha^i \cdot \sum_l f^*_l T_l(f_l^*)} 
\\ 
&= \frac{\max_{i \in \cN} \alpha^i}{\min_{i \in \cN} \alpha^i} \cdot \overline{PoS}.
\end{align*}
The lower bound follows similarly by switching {\em min} with {\em max}.
} 
\end{IEEEproof}
As a result of Proposition \ref{PoS_upper_bounded}, we get the following corollary.
\begin{corollary}
In the game defined in Section~\ref{sec:model}, with two users or with $N$ identical users, 
with homogeneous costs and with a weighted social cost,
the {\em PoS} is upper bounded by,
$$PoS \leq \frac{\max_{i \in \cN} \alpha^i}{\min_{i \in \cN} \alpha^i}.$$ 
\end{corollary}


\section{Coping with Heterogeneity}\label{sec:PoH}
\subsection{Unbounded PoS}
It is of interest to consider the difference between the {\em PoS}
and {\em PoA} also within the wider class of standard functions.
The following result establishes that there are instances where such a difference does not exist and moreover, both the {\em PoS} and the {\em PoA} may assume arbitrarily 
(and identically) large values. 
Consequently, in such cases, the $N$-player bargaining problem is not {\em essential}.
\begin{theorem}
\label{PoS_large}
In the game defined in Section~\ref{sec:model}, with standard costs, the corresponding $N$-player bargaining problem may not be {\em essential}, i.e., the NBS may coincide with the NEP.
Moreover,
the {\em Price of Selfishness} (hence, also the {\em Price of Anarchy}) can
be arbitrarily large.
\end{theorem}
\begin{IEEEproof}
We establish the claim through the following example.
Consider a network with two users and two parallel links and let the total demands be $r^1 = r^2 =0.5$.
The costs of the users are defined as follows:
\begin{align}\label{eq:NE_costs}
J^1 &= f^1_1 \cdot f_1 + 2 \cdot f_2^1 \cdot ( f_2 + 1)
 \\ \nonumber J^2 &= \frac{f_1^2}{1+\epsilon - f_1}+\frac{2}{\epsilon^2} \cdot f_2^2 \cdot ( f_2 + 1)
\end{align}
where $0 < \epsilon < 0.1$. It can be verified that, at
the NEP, both users ship all of their flow through link $1$. Therefore, their costs at the NEP are given by $\hat{J^1} = 0.5$
and $\hat{J^2} = \frac{0.5}{\epsilon}$.

We first need to verify that it is not profitable for any of the users to flow on the bottom link at the NBS. This has to hold for any bargained probability vector $\bf{\tilde{p}}$. 

Consider $M$ feasible cost vectors $J({\bf{ \tilde{f}(m)}}), m = 1 \ldots M$ and denote the flows of the users $1$ and $2$ on the bottom link as, respectively, $x_m$ and $y_m$.


The individual costs of the users at the cost vectors $J^i({\bf \tilde{f}(m)}), m = 1 \ldots M$,$~i=1,2$ are equal to:
\begin{align}
\label{strat_cost}
J^1(\mathbf{\tilde{f}(m)}) &= (\frac{1}{2} - x_m)(1-x_m-y_m) + 2x_m \cdot (x_m+y_m + 1) \\ \nonumber
J^2(\mathbf{\tilde{f}(m)}) &= \frac{\frac{1}{2}-y_m}{\epsilon + x_m+y_m} + \frac{2y_m \cdot (x_m+y_m+1)}{\epsilon^2}.
\end{align}
We have to show that none of the players use the bottom link at the NBS, i.e. the NEP coincides with the NBS. 
Due to Axiom $N1$ (Individual Rationality), the following should hold for $i = 1,2$. 
\begin{equation}\label{eq:expected_prof_loss}
\sum_{m=1}^M \left[ \tilde{p}_m \cdot (J^i({\bf \tilde{f}(m)}) - \hat{J}^i ) \right] \leq 0.
\end{equation}

From (\ref{strat_cost}), the cost of the first user can be written as
\begin{equation}\label{eq:profit_loss_0}
J^1(\mathbf{\tilde{f}(m)}) = \frac{1}{2} + \frac{1}{2} x_m - \frac{1}{2} y_m + 3(x_m)^2 + 3x_m \cdot y_m.
\end{equation}
The following inequality now follows from (\ref{eq:expected_prof_loss}) and (\ref{eq:profit_loss_0}), for all $j = 1,2, \ldots, M$:
\begin{gather}
\label{profit_loss_1}
\sum_{m \backslash j} \left[\tilde{p}_m \cdot \left( \frac{1}{2} + \frac{1}{2}x_m - \frac{1}{2}y_m + 3(x_m)^2 + 3x_m \cdot y_m - \hat{J}^1 \right) \right]\\ 
\leq \tilde{p}_j \cdot \left(  \hat{J}^1 - \frac{1}{2} - \frac{1}{2}x_j + \frac{1}{2}y_j - 3(x_j)^2 - 3x_j \cdot y_j \right) \nonumber
\end{gather}
where $\hat{J}^1 = \frac{1}{2}$. In the same way, we get for the second user\\
$\forall j = 1,\ldots,M$
\begin{gather*}
\label{profit_loss_2}
\sum_{m \backslash j} \left[\tilde{p}_m \cdot \left( \hat{J}^2 - \frac{\frac{1}{2}-y_m}{\epsilon + x_m+y_m} - \frac{2y_m \cdot (x_m+y_m+1)}{\epsilon^2} \right) \right] \\
\geq \tilde{p}_j \cdot \left( \frac{\frac{1}{2}-y_j}{\epsilon + x_j+y_j} + \frac{2y_j \cdot (x_j+y_j+1)}{\epsilon^2} - \hat{J}^2 \right)  \nonumber
\end{gather*}
where $\hat{J}^2 = \frac{0.5}{\epsilon}$. 
Thus,
we obtain for the second user \\
$\forall j = 1,\ldots,M$
\begin{gather*}
\sum_{m \backslash j} \left[ \tilde{p}_m \left( \frac{\frac{1}{2}(x_m + y_m) + \epsilon \cdot y_m}{\epsilon \cdot (\epsilon + x_m+y_m)} -  \frac{2y_m \cdot (x_m+y_m+1)}{\epsilon^2} \right) \right] \\ \nonumber
\geq \tilde{p}_j \left( -\frac{\frac{1}{2}(x_j + y_j) + \epsilon \cdot y_j}{\epsilon \cdot (\epsilon + x_j+y_j)} +  \frac{2y_j \cdot (x_j+y_j+1)}{\epsilon^2} \right).
\end{gather*}
Thus, 
it holds that$~\forall j = 1,\ldots,M$:
\begin{gather*}
\sum_{m \backslash j} \left[ \tilde{p}_m \left( \frac{\frac{1}{2}(x_m + y_m) + \epsilon \cdot y_m - 2y_m \cdot (x_m+y_m+1)}{\epsilon^2} \right) \right] \\
\geq \tilde{p}_j \left( \frac{-\frac{1}{2}(x_j + y_j) - \epsilon \cdot y_j + 2y_j \cdot (x_j+y_j+1)}{\epsilon^2} \right). \nonumber   
\end{gather*}
Simplifying the above, we get $~\forall j = 1,\ldots,M$:
\begin{gather}
\sum_{m \backslash j} \left[ \tilde{p}_m \left(\frac{1}{2}(x_m - y_m) - y_m(1-\epsilon) - 2(y_m)^2 - 2x_m \cdot y_m \right) \right] \label{profit_loss_6} \\
\geq \tilde{p}_j \left( -\frac{1}{2}(x_j - y_j) + y_j(1-\epsilon) + 2(y_j)^2 + 2x_j \cdot y_j \right).   \nonumber
\end{gather}
By substracting (\ref{profit_loss_1}) from (\ref{profit_loss_6}) we obtain:
\begin{gather} 
\label{contra_1}
\sum_{m \backslash j} \left[ \tilde{p}_m \left(- y_m(1-\epsilon) - 2(y_m)^2 - 5x_m \cdot y_m - 3(x_m)^2\right) \right] \\
\geq \tilde{p}_j \left(y_j(1-\epsilon) + 2(y_j)^2 + 5x_j \cdot y_j + 3(x_j)^2 \right) ~\forall j = 1,\ldots,M.   \nonumber
\end{gather}

Since, by definition, $\tilde{p}_m > 0$ for all $m = 1, 2,\ldots, M$,
(\ref{contra_1}) can hold only if $x_m = y_m = 0$, for all $m = 1, 2,\ldots, M$.
Hence, we establish that at the Nash Bargaining Scheme, the players will only use the top link. This also shows that the Nash Equilibrium is Pareto optimal. Therefore, the NBS in this case coincides with the NEP (i.e. the {\em PoS} equals the {\em PoA}).

Consider now the following (feasible) flow profile: $f_1^1 =0$, $f_2^1 = 0.5$, $f_1^2 =0.5$, $f_2^2 = 0$. The corresponding social cost is
equal to $1.5 + \frac{0.5}{0.5+\epsilon}$, i.e., smaller than $2.5$. Therefore, the social optimum is no more than $2.5$.
Thus:
\begin{equation}
PoS = PoA \geq  \frac{0.5+\frac{0.5}{\epsilon}}{2.5} > \frac{0.2}{\epsilon}.
\end{equation}
Hence, the {\em PoS} and the {\em PoA} can be made arbitrarily large by choosing a sufficiently small $\epsilon$.
\end{IEEEproof}
More generally, it is interesting to note that, due to the axiom of Individual Rationality, whenever the Nash Equilibrium is Pareto optimal (as in the above example), 
the {\em Price of Selfishness} equals the {\em Price of Anarchy}. 

\subsection{The Price of Heterogeneity}
Theorem~\ref{PoS_large} 
is not surprising. Indeed,
with non-homogeneous costs, each user may be trying to optimize completely different performance objectives. Hence what might be ``good'' for one might be ``bad''
for the other, in which case there is little hope for bargaining. Moreover, with heterogeneous objectives, 
a social objective that is some simple combination of the individual cost functions, e.g. their sum, may be artificial. 
This implies that the Prices of Anarchy and Selfishness,
which are based on the definition of such a social cost, may be inappropriate. 
Hence, for the heterogeneous case 
we need to look for an alternative concept for benchmarking the
deterioration of performance due to the competition among players.  

To that end, consider again the example in the proof of Theorem~\ref{PoS_large}. To achieve social optimum, user $1$ would need to sacrifice its performance
and also use link $2$. However, that link is also costlier for user $1$ (albeit not to the extent it is for user $2$), hence,
both at an NEP and at an NBS, user $1$ would stick with link $1$.
Now, suppose we optimally routed all of the traffic (of both users), but {\em considering as the target, the cost function of user $2$}. In that case, the (socially) optimal solution would coincide with the NBS (and the NEP), i.e., all traffic routed over link $1$, with an optimal (arbitrarily large) cost value of $\frac{1}{\epsilon}$. Hence, the problem here is not due to selfish behavior but
rather due to the poor performance of the network, as seen from the perspective of user $2$. 
The above discussion suggests that, with heterogeneous users, the deterioration of performance 
in the game scenario should
be measured through the following question: how much might the performance of a user deteriorate, due to the selfish behavior of the other users,
with respect to the case where {\em all the traffic} would be optimally controlled {\em according to its own cost function}. This figure measures
the price that a user pays for the plurality of performance objectives. 

\begin{definition} \label{def:cost_perceived}
For a user $i$ with a (standard) cost function $J^i$, the {\em system cost function perceived by user $i$}, 
denoted by $J_{sys}^i$, is the cost function obtained 
when applying $J^i$ to the whole of the system traffic $R= \sum_{j \in \cN} r^j$.
\end{definition}
For example, consider a user $i$ that attempts to minimize its delay, where $T_l (\fl)$ stands for the link delay. 
Then, \\
$J^i = \sum_{l \in \cL} f_l^i \cdot T_l (f_l)$, whereas $J_{sys}^i =  \sum_{l \in \cL} f_l \cdot T_l (f_l)$.

However, $J^i$ and $J_{sys}^i$ are not comparable. In the above example, $J^i$ 
is the {\em total} delay experienced by a volume of traffic of size $r^i$, whereas $J_{sys}^i$ 
is the {\em total} delay experienced by a (larger) volume of traffic of size $R$.
Rather, the figures that should be compared are the respective performances {\em of each unit of traffic}, i.e., we should {\em normalize}
$J^i$ and $J_{sys}^i$ by the respective volumes of traffic $r^i$ and $R$. Indeed, in the above example, this would compare
between the respective {\em average delays}, namely $\frac{1}{r^i} \cdot \sum_{l \in \cL} f_l^i \cdot T_l (f_l)$ and
$\frac{1}{R} \cdot \sum_{l \in \cL} f_l \cdot T_l (f_l)$. We thus define:
\begin{definition} \label{def:norm_cost}
The {\em normalized cost} of a user $i$ is the ratio between its cost $J^i$ and its size $r^i$, namely: $\frac{1}{r^i} \cdot J^i$.
Similarly, the {\em normalized system cost perceived by user $i$} is the ratio between $J_{sys}^i$ and the
total size of the system $R$, namely: \\
$\frac{1}{R} \cdot J_{sys}^i$.
\end{definition}
Note that Definition \ref{def:cost_perceived} implicitly assumes that there is a way to define the system cost of a game through the cost function of a specific user. 
While this is a valid assumption in many classes of games, there are cases, such as zero-sum games, where it is not.
Moreover, Definition \ref{def:norm_cost} is only applicable to games where there exists a way to measure the size of the players, e.g, number of items, budget. In the realm of routing games, the size of a player refers to its flow demand.
We are now ready to define our proposed concept for quantifying the degradation of performance in the heterogeneous case.

\begin{definition}
The {\em Price of Heterogeneity of a user $i \in \cN$ $(PoH^i)$} is the ratio between the normalized cost experienced by that user at a 
(worst, if many) Nash Equilibrium and the optimal value of the normalized system cost perceived by that user, i.e.,
\begin{equation}
\label{eq:poh_i}
PoH^i = \frac{\frac{1}{r^i}\cdot\hat{J}^i}{\frac{1}{R} J_{sys}^{i*}},
\end{equation}
where 
$J_{sys}^{i*}$ is the minimum value of $J_{sys}^{i}$, namely \\
$J_{sys}^{i*} = \min_{\bff \in \bf {F}} J_{sys}^{i}$.
\end{definition} 
Similarly, the {\em Price of Heterogeneity (PoH)} is the worst value of the $PoH^i$, i.e.:
\begin{equation}
\label{eq:poh}
PoH = \max_{i \in \cN} PoH^i.
\end{equation}
We proceed to establish an upper-bound on the {\em PoH} for the class of routing (load balancing) games specified in Section~\ref{sec:model},
considering the general (and, potentially ``highly heterogeneous'') class of standard functions. 

\begin{theorem}
\label{PoS_wc}
In the game defined in Section~\ref{sec:model}, with standard costs, the following hold:
\begin{itemize}
\item The Price of Heterogeneity of a user $i \in \cN$, i.e., $PoH^i$, is upper-bounded by
$PoH^i \leq \frac{R}{r^i}$. 
\item Let $r = \min_{i \in \cN} r^i$. Then, the Price of Heterogeneity, i.e., $PoH$, is upper-bounded by
$PoH \leq \frac{R}{r}$. 
\end{itemize}
\end{theorem}
\begin{IEEEproof}\footnote{We note that this proof can be easily adapted to obtain the following upper-bound for the Price of 
Anarchy for users with homogeneous costs:
$PoA \leq N$. However, that bound has already been established in~\cite{harks_09}.}
Consider a user $i \in \cN$. Denote by $\hat{f}_l^{-i}$ the aggregate
flow of the other users at the NEP on link $l \in \cL$, i.e., $\hat{f}_l^{-i} = \sum_{j \in \cN, j \neq i} \hat{f}_l^j$, and
denote by $R^{-i}$ their aggregated demand, i.e., $R^{-i} = \sum_{j \in \cN, j \neq i} r^j$.
We will now describe a ``benchmark'' strategy, $\bf \bar{f^i}$, which is a feasible routing strategy that
user $i$ is able to use given the strategies of the
other users as provided by the flows $\hat{f}_l^{-i}$; we point
out that this is not necessarily the best reply of user $i$ to the $\hat{f}_l^{-i}$s. 

Consider the problem of minimizing $J_{sys}^i$, i.e., the problem of optimally routing the total demand $r^i+R^{-i}$ according to the cost function of user $i$, i.e., $J^i$. Also,
denote $R^{-i}(0) = R^{-i}$ and $\cL_0 = \cL$.
Denote by $f_l^{i*}(0)$  the corresponding optimal flow on link $l \in \cL$. We thus have:
\begin{equation}\label{worst_opt}
J_{sys}^{i*} = \sum_{l \in \cL_0} J^{i}_l(f_l^{i*}(0),f_l^{i*}(0)).
\end{equation}

Denote by $\cL_1 \subseteq \cal{L}$ the set of links for which, $\forall l\in \cL_1$, $\hat{f}_l^{-i} \leq f_l^{i*}(0)$. Clearly, this set is nonempty.
Denote $R^{-i}(1) \triangleq R^{-i}(0) - \sum_{l \notin \cL_1} \hat{f}_l^{-i}$, i.e., $R^{-i}(1)$ is the total amount of flow that the other users ship on the links 
where $\hat{f}_l^{-i} \leq f_l^{i*}(0)$. 
Consider now the problem of optimally routing the total demand $r^i+R^{-i}(1)$, according to the cost function $J^i$,
{\em solely} over the links in $\cL_1$.
Denote by $f_l^{i*}(1)$  the corresponding optimal flow on link $l \in \cL_1$.  

Continuing inductively, denote by $\cL_{k+1} \subseteq \cL_k$ the set of links for which, $\forall l\in \cL_{k+1}$, $\hat{f}_l^{-i} \leq f_l^{i*}(k)$; it is clear that this
set is nonempty. Denote $R^{-i}(k+1) \triangleq R^{-i}(k) - \sum_{l \in \cL_{k} \setminus\cL_{k+1}} \hat{f}_l^{-i}$, i.e., $R^{-i}(k+1)$ is the total amount of flow that the other
users ship on the links 
where $\hat{f}_l^{-i} \leq f_l^{i*}(k)$.
The process continues until, necessarily, for some $K$, ${\cal{L}}_K =$ ${\cal{L}}_{K+1}$.

The benchmark strategy is defined as follows: in each link $l \in \cL_K$ (for which, by construction, $\hat{f}_l^{-i} \leq f_l^{i*}(K)$), user $i$
ships $\bar{f}_l^i = f_l^{i*}(K) - \hat{f}_l^{-i}$, whereas in other links it does not ship any flow. 
Clearly, this is a feasible strategy
for $i$, since $\bar{f}_l^i \geq 0$ for all $l \in \cL$ and 
\begin{equation}\sum_{l \in \cL} \bar{f}_l^i = \sum_{l \in \cL_K} (f_l^{i*}(K) - \hat{f}_l^{-i}) = (R^{-i}(K)+r^i)-R^{-i}(K)= r^i.
\end{equation}
We proceed with the following lemma.
\begin{lemma}\label{worst_case}
For all $k$, $1 \leq k \leq K$, 
\begin{equation}
\sum_{l \in \cL_k} J^i_l(f_l^{i*}(k), f_l^{i*}(k)) \leq \sum_{l \in \cL_{k-1}} J^i_l(f_l^{i*}(k-1), f_l^{i*}(k-1)).
\end{equation}
\end{lemma}
\begin{IEEEproof}
\ifthenelse{\equal{\Context}{PAPER}}
{
See~\cite{BO11}.
}
{
By construction, for $1 \leq k \leq K$:
\begin{equation}
\sum_{l \in \cL_k} f_l^{i*}(k) = r^i + R^{-i}(k) = r^i + R^{-i}(k-1) - \sum_{l \in \cL_{k-1} \setminus \cL_k} \hat{f}_l^{{-i}}.
\end{equation}
Furthermore, by construction:
\begin{equation}
\sum_{l \in \cL_{k-1} \setminus \cL_k} \hat{f}_l^{{-i}}\geq \sum_{l \in \cL_{k-1} \setminus \cL_k} f_l^{i*}(k-1).
\end{equation}
Therefore:
\begin{align} \label{eq:worst_ineq}
\sum_{l \in \cL_k} f_l^{i*}(k) &= r^i + R^{-i}(k-1) - \sum_{l \in \cL_{k-1} \setminus \cL_k} \hat{f}_l^{{-i}} \\ \nonumber
&\leq r^i + R^{-i}(k-1) - \sum_{l \in \cL_{k-1} \setminus \cL_k} f_l^{i*}(k-1) \\ \nonumber 
&= \sum_{l \in \cL_k} f_l^{i*}(k-1).
\end{align}

Now, since $(f_l^{i*})(k)_{l \in \cL_{k}}$ is an {\em optimal} flow (for the function $J^i$) for routing a total amount of $\sum_{l \in \cL_k} f_l^{i*}(k)$
over the links of $\cL_k $, while $(f_l^{i*})(k-1)_{l\in\cL_{k}}$ is some flow for routing a {\em larger} total amount (in view of~(\ref{eq:worst_ineq}))
over {\em the same set of links}, due to the monotonicity assumption {\em S3} on standard cost functions, it follows that
\begin{align*}
\sum_{l \in \cL_k} J^i_l(f_l^{i*}(k), f_l^{i*}(k)) 
&\leq \sum_{l \in \cL_{k}} J^i_l(f_l^{i*}(k-1), f_l^{i*}(k-1))
\\ 
&\leq \sum_{l \in \cL_{k-1}} J^i_l(f_l^{i*}(k-1), f_l^{i*}(k-1))
.
\end{align*}
} 
\end{IEEEproof}

Applying the above Lemma inductively on $k$, we get:
\begin{equation}
\sum_{l \in \cL_K} J^i_l(f_l^{i*}(K), f_l^{i*}(K)) \leq \sum_{l \in \cL_0} J^i_l(f_l^{i*}(0), f_l^{i*}(0)) = J_{sys}^{i*},
\end{equation}
where the last identity follows from (\ref{worst_opt}). 
On the other hand, the cost $\bar{J}^i$ that user $i$ experiences when employing the benchmark strategy against the strategies of the other users $j \neq i$ satisfies:
\begin{align}\label{worst_last}
\bar{J}^i &= \sum_{l \in \cL_K} J^i_l(\bar{f}_l^{i*}, \bar{f}_l^{i*}) 
\\ \nonumber
&= \sum_{l \in \cL_K} J^i_l(f_l^{i*}(K) - \hat{f}_l^{-i}, f_l^{i*}(K) - \hat{f}_l^{-i})
\\ \nonumber 
&\leq \sum_{l \in \cL_K} J^i_l(f_l^{i*}(K), f_l^{i*}(K)),
\end{align}
where the inequality in (\ref{worst_last}) is due to the monotonicity assumption {\em S3} on standard cost functions.
Therefore $\bar{J^i} \leq J_{sys}^{i*}$.
By definition, at the Nash Equilibrium, user $i$ employs a best-reply strategy for the given strategies of the other players,
hence its cost at the Nash Equilibrium, $\hat{J^i}$, is no larger than when employing the benchmark strategy,
i.e., $\hat{J^i} \leq \bar{J^i}$, hence, $\hat{J^i} \leq J_{sys}^{i*}$. 
Normalizing the costs, we have:
\begin{equation}\label{eq:ratio}
PoH^i = \frac{\frac{1}{r^i}\cdot\hat{J}^i}{\frac{1}{R} J_{sys}^{i*}}  = \frac{R}{r^i} \cdot \frac{\hat{J}^i}{J_{sys}^{i*}} \leq \frac{R}{r^i} 
\end{equation}
hence establishing the first part of the theorem. The second part follows immediately.
\end{IEEEproof}

The Price of Heterogeneity of a user $i$, $PoH^i$, could be interpreted as the {\em PoA} 
``as seen by that user''. Per Theorem~\ref{PoS_wc}, 
that value cannot be worse than the reciprocal of the user's relative size in the system. A practical implication of the theorem is that,
in a heterogeneous environment, 
``you'd better be big''.\footnote{In view of the {\em bargaining paradox}~\cite{rat_behav}, the above is only in the worst- 
case sense considered by the $PoH$.} In a way, the theorem provides additional incentives to cooperate 
with other users that contemplate the same performance objectives, even in the 
presence of other heterogeneous users. 
Indeed, 
from the worst-case perspective considered by the {\em PoH}, Theorem~\ref{PoS_wc} indicates that, through bargaining, a homogeneous subset of users can obtain a strategy profile that would effectively make them behave as a single user with an aggregated traffic demand, hence decreasing the upper bound on the {\em PoH}.

Another practical implication from the derived upper bound on the {\em PoH} is that, 
from a system perspective, users (or groups of homogeneous users) are preferred to be identical in demand.
This seemingly contradicts \cite{KorilisLO97}, which considers a scenario where a network administrator is able to send its own demand through the network and aims at optimizing the overall
system performance. 
It is concluded in \cite{KorilisLO97} that,
when the demands of all users are equal, it is hard for a network administrator to enforce the system optimum. 
Our result adds to \cite{KorilisLO97} and illustrates that, in a scenario without direct interference from a network administrator, the system may perform better if users have identical demands. 

\section{Conclusion}\label{sec:conclusion}

We investigated the added value of bargaining among players in a communication
network. As a new figure of merit for cooperative games, the {\em Price of Selfishness}
was introduced and under the case of homogeneous costs, the NBS guarantees an improvement in performance for all users and for the system. Moreover, for certain cases,
the NBS was shown to be equal to the optimal (social) solution of the system.
It remains an open question how to tighten the bounds on the {\em PoS} for the general case with $N$-users. 
We also considered the case of non-homogeneous costs, for which we proposed the
{\em Price of Heterogeneity} as
an appropriate extension of the {\em Price of Anarchy} and established an upper bound on the {\em PoH} under quite general conditions.

Our study focused on load balancing (routing) among servers (links), and
furthermore, it considered a specific solution concept, namely the NBS.
Yet, we believe that it provides useful insight into the potential merit of
adopting bargaining schemes in networking games.
For example, having {\em PoS} $\equiv 1$ but potentially {\em PoA} $>> 1$ in certain homogeneous settings, together with a potentially unbounded {\em PoS} in non-homogeneous
settings, suggests a design guideline that attempts
to separate among homogeneous groups of users (e.g., ``highly delay-sensitive'',
``less delay-sensitive but highly sensitive to packet loss'', etc.) so that each group would share its own network resources.
Another important issue is the ability, or willingness, of the users to bargain.
The essential bargaining solution concept implicitly assumes that, as long as Axiom {\em N1} is met, the players would accept the solution. Yet, some users might
not be able to do so, either due to technical reasons (e.g., inability to
communicate) or to other reasons, such as administrative or legal
constraints. Such a user, even if ``homogeneous''  in terms of its cost
function, may prevent achieving social optimality through bargaining.
Therefore, another design guideline would be to try to separate between cooperative
users (i.e., that can engage in bargaining) and those that do
not cooperate. Furthermore, if such isolations are not possible and we need to confront a
heterogeneous scenario, our bound on the {\em PoH} suggests that
homogeneous groups of users would benefit from bargaining a joint strategy.


While we advocated the choice of the NBS, other solution concepts of
cooperative games should be considered, as they could
better fit some of the networking scenarios. For example, agents may have different bargaining powers, hence asymmetric bargaining schemes~\cite{Myerson} might be called for. 
In addition, the NBS contemplates two scenarios, namely a grand
coalition versus a ``disagreement point''.
Yet, partial coalitions should also be taken into account, e.g., due to the inability of
some users to engage in bargaining. 
Finally, we aim to consider more complex topologies, which correspond to a larger range of networking scenarios.  Investigating the added value of bargaining in such
contexts is thus another important area for future work.


\bibliographystyle{IEEEtran}
\bibliography{IEEEabrv,UN_bibfile_new_v3}

\end{document}